\documentclass{PoS}
\usepackage{wrapfig}
\def\GeV{{\rm GeV}}
\def\TeV{{\rm TeV}}

\title{PDFs and Top Physics}

\ShortTitle{PDFs}

\author{\speaker{R.S.~Thorne} \\
        Department of Physics and Astronomy, \\
        University College London, WC1E 6BT, UK\\
        E-mail: \email{robert.thorne@ucl.ac.uk}}

\abstract{I present the results from the recent PDF4LHC study,
and the resulting new recommendation for combining PDFs sets for LHC
calculations. In order to put this into context I summarise
continuing developments in PDFs. This includes improvements and recent
updates of particular PDF sets due to theory improvements and a variety of
new data sets, including most of the up-to-date LHC data.
I will emphasise particular issues relevant for top physics.}

\FullConference{8th International Workshop on Top Quark Physics\\
                 14-18 September, 2015\\
                 Ischia, Italy}

\begin{document}

\section{Recent PDF Updates - effect and treatment of LHC data}

\begin{wrapfigure}{r}{0.6\columnwidth}
\vspace{-0.4cm}
\centerline{\includegraphics[width=0.62\textwidth]{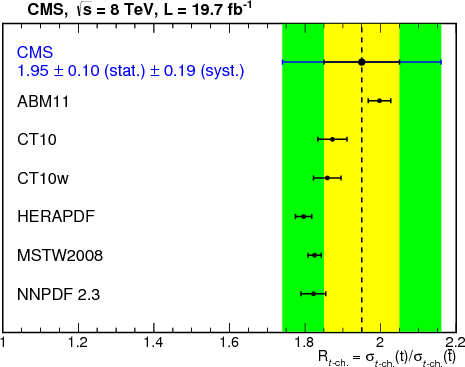}}
\vspace{-0.3cm}
\caption{The CMS measurement of the $t/\bar t$ ratio in $t$-channel production, figure from 
\cite{Khachatryan:2014iya}.}
\vspace{-0.3cm}
\label{Fig2}
\end{wrapfigure}

Each group has produced updates including new data, often including data from 
the LHC.  
The recent analysis from the ABM group, ABM12~\cite{Alekhin:2013nda},
now includes more HERA cross-section data, and vector 
boson production data from ATLAS, CMS and LHCb. 
The PDF sets are determined together with 
$\alpha_S$, whose value  comes out to be  $\alpha_S(m_Z^2)=0.1132$ 
at NNLO. Top quark pair production data from the LHC is investigated, but 
not included in the 
default PDFs. Its inclusion tend to raise the high-$x$ gluon and 
$\alpha_S(m_Z^2)$ a little,
the precise details depending on the top quark mass (and mass renormalization 
scheme) used.  ABM PDF sets currently give the best fit to the ratio of 
$t$-channel single top to single anti-top production 
\cite{Khachatryan:2014iya,Aad:2014fwa}, as seen in Fig.~\ref{Fig2}, which is 
a constraint on $u/d$. 

\begin{figure}
\vspace{-0.2cm}
\centerline{\includegraphics[width=0.5\textwidth]{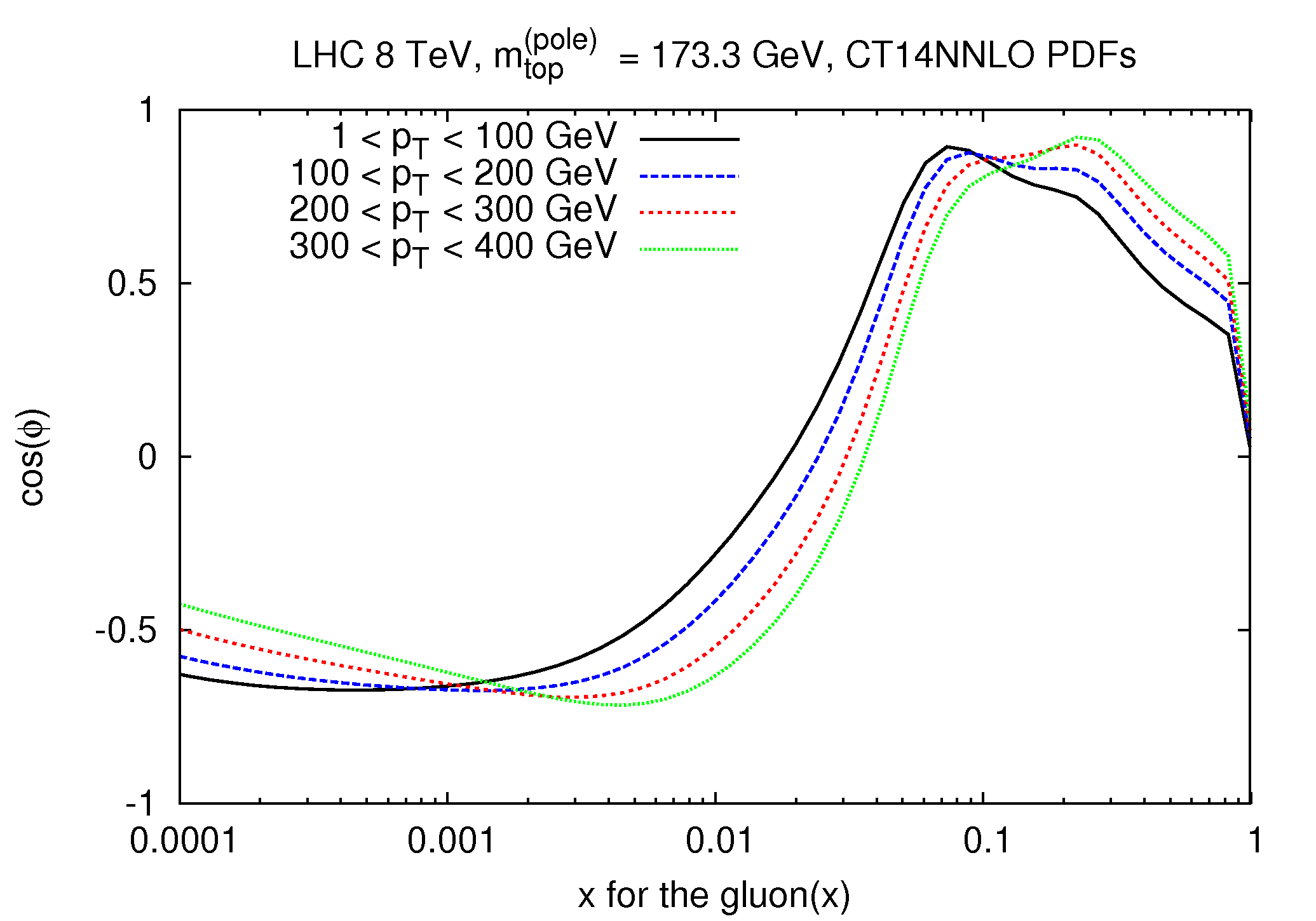}
\includegraphics[width=0.5\textwidth]{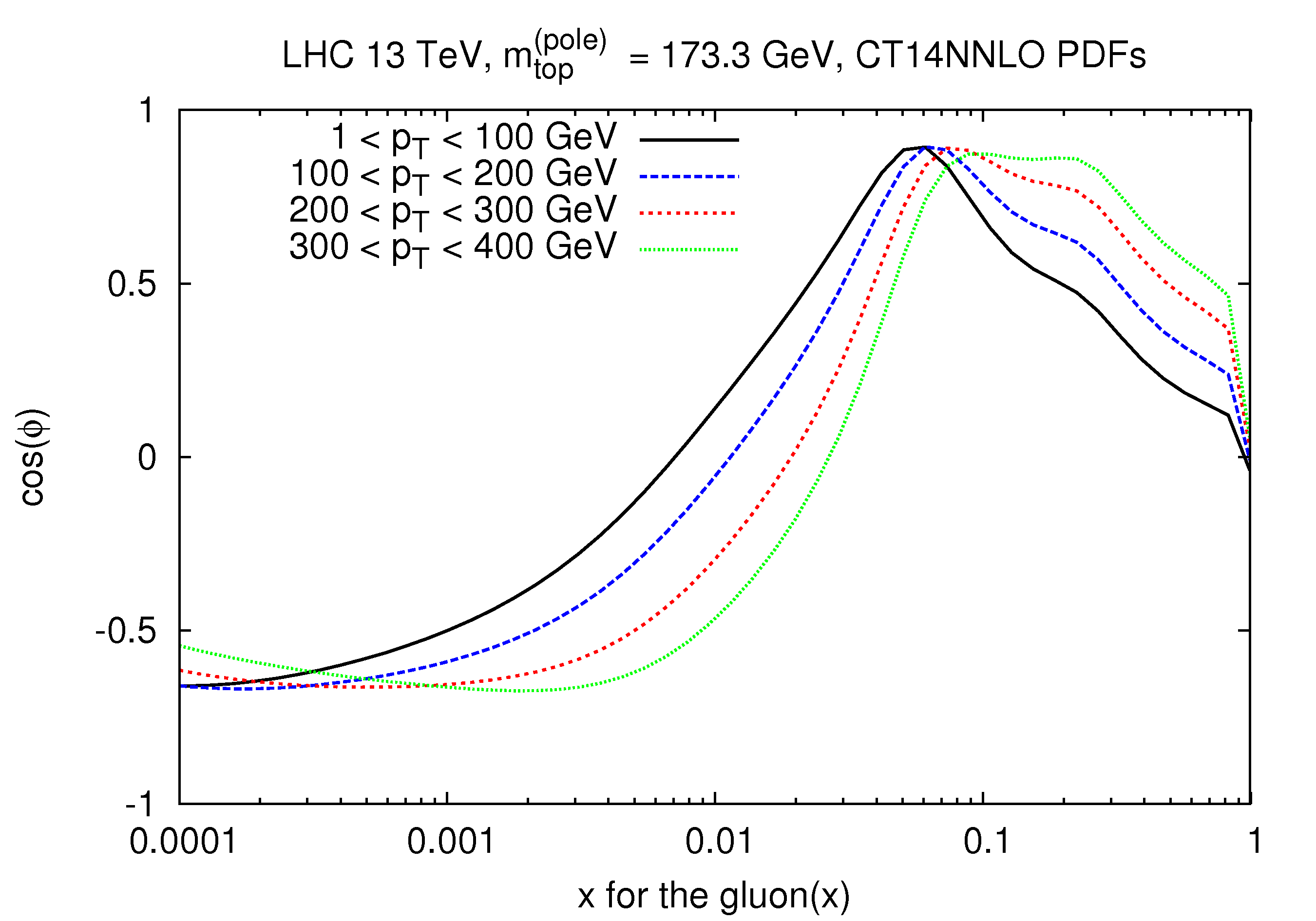}}
\vspace{-0.1cm}
\caption{The correlation between top pair production in different b$p_T$ bins 
and the gluon, figures from \cite{Dulat:2015mca}. }
\vspace{-0.3cm}
\label{Fig3}
\end{figure}

The CT14 PDF sets~\cite{Dulat:2015mca} have been made recently available 
at NLO, NNLO, and also at LO.
These sets include a variety 
of LHC data sets as well as the most recent D0 data on electron charge 
asymmetry. The PDFs also use an updated parametrization based on 
Bernstein polynomials which peak at a specific $x$. LHC inclusive jet data
are included at NLO and also in the NNLO fit.
The main change in the PDFs as
compared to CT10 is a softer high-$x$ gluon, a smaller strange quark (partially due to 
correction of the charged current DIS cross section code) and the 
details of the flavour decomposition, e.g. $\bar u /\bar d$ and the high-$x$ 
valence quarks. CT14 does not fit top quark production data but does 
make comparisons, e.g. the correlation of top pair production with the gluon 
are shown in Fig.{\ref{Fig3}}.

\begin{wrapfigure}{r}{0.57\columnwidth}
\vspace{-0.6cm}
\centerline{\includegraphics[width=0.72\textwidth]{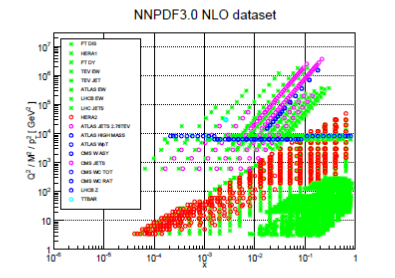}}
\vspace{-0.1cm}
\caption{The data included in the NNPDF3.0 analysis, figure from \cite{Ball:2014uwa}. }
\vspace{-0.6cm}
\label{Fig4}
\end{wrapfigure}

The NNPDF3.0 PDFs \cite{Ball:2014uwa} are the recent major
update within the NNPDF framework. As new data they include HERA inclusive
structure function Run II data from H1 and ZEUS (before their combination),
more  recent ATLAS, CMS and LHCb data on gauge boson production and
inclusive jets, and  $W+$charm  and top quark pair production.
A subset of jet data is included at NNLO using an approximate NNLO treatment.
The full set of data fit is illustrated in Fig.~\ref{Fig4}. 
The NNPDF3.0 fitting procedure has been tuned using a 
closure test, i.e, by generating pseudo-data based on an assumed
underlying set of PDFs. One verifies in this case that the output of the
fitting procedure is consistent with the a priori known answer. 
As a by-product, one can investigate directly the origin
of PDF uncertainties. The minimization has been
optimized based on the closure test.  
The NNPDF3.0 PDFs display moderate changes in comparison to NNPDF2.3:
specifically somewhat smaller  uncertainties and a 
noticeable change in the gluon-gluon luminosity which is mainly due 
to the change in methodology. 

\begin{figure}
\vspace{-0.0cm}
\includegraphics[width=0.5\textwidth]{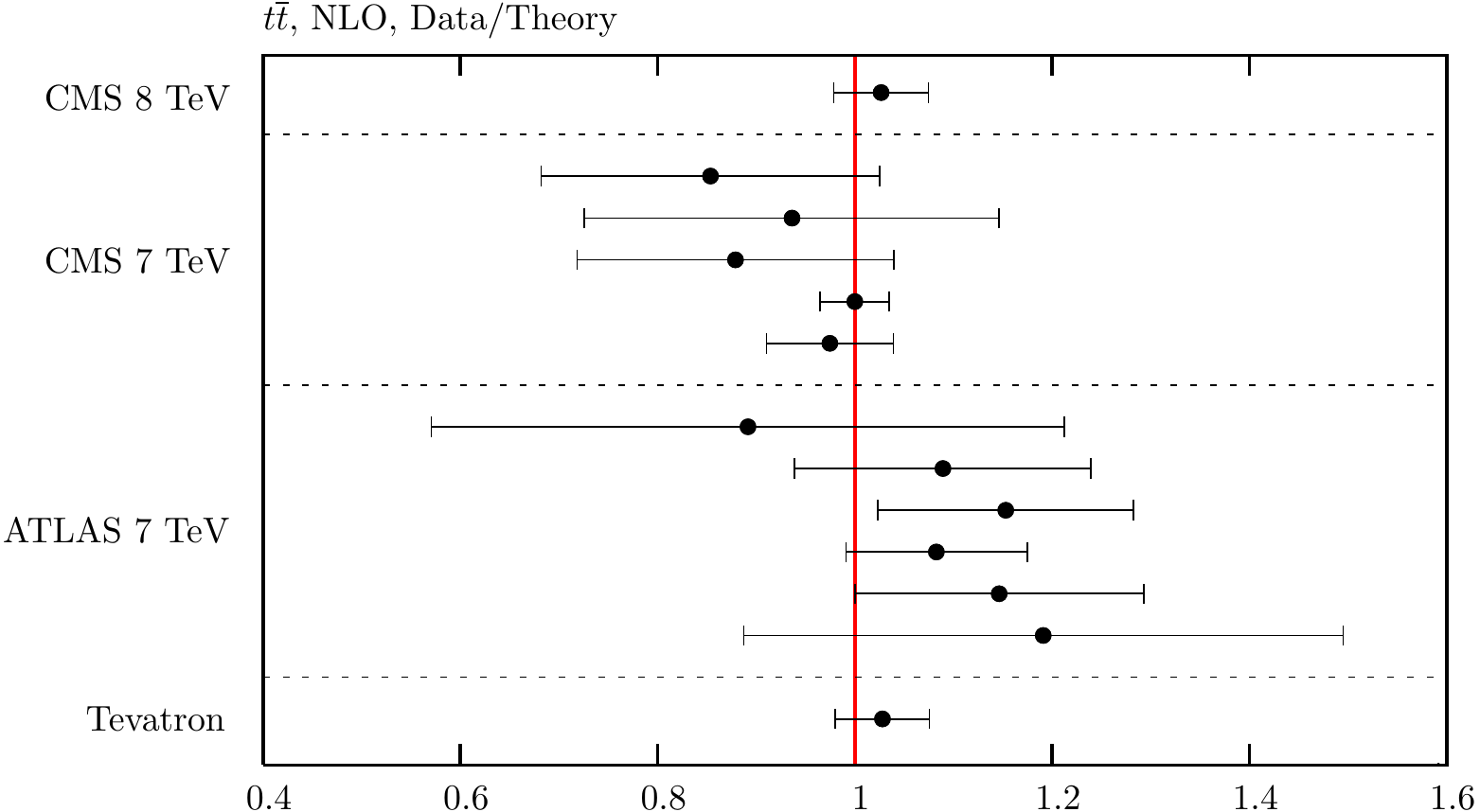}
\vspace{0.2cm}
\includegraphics[width=0.5\textwidth]{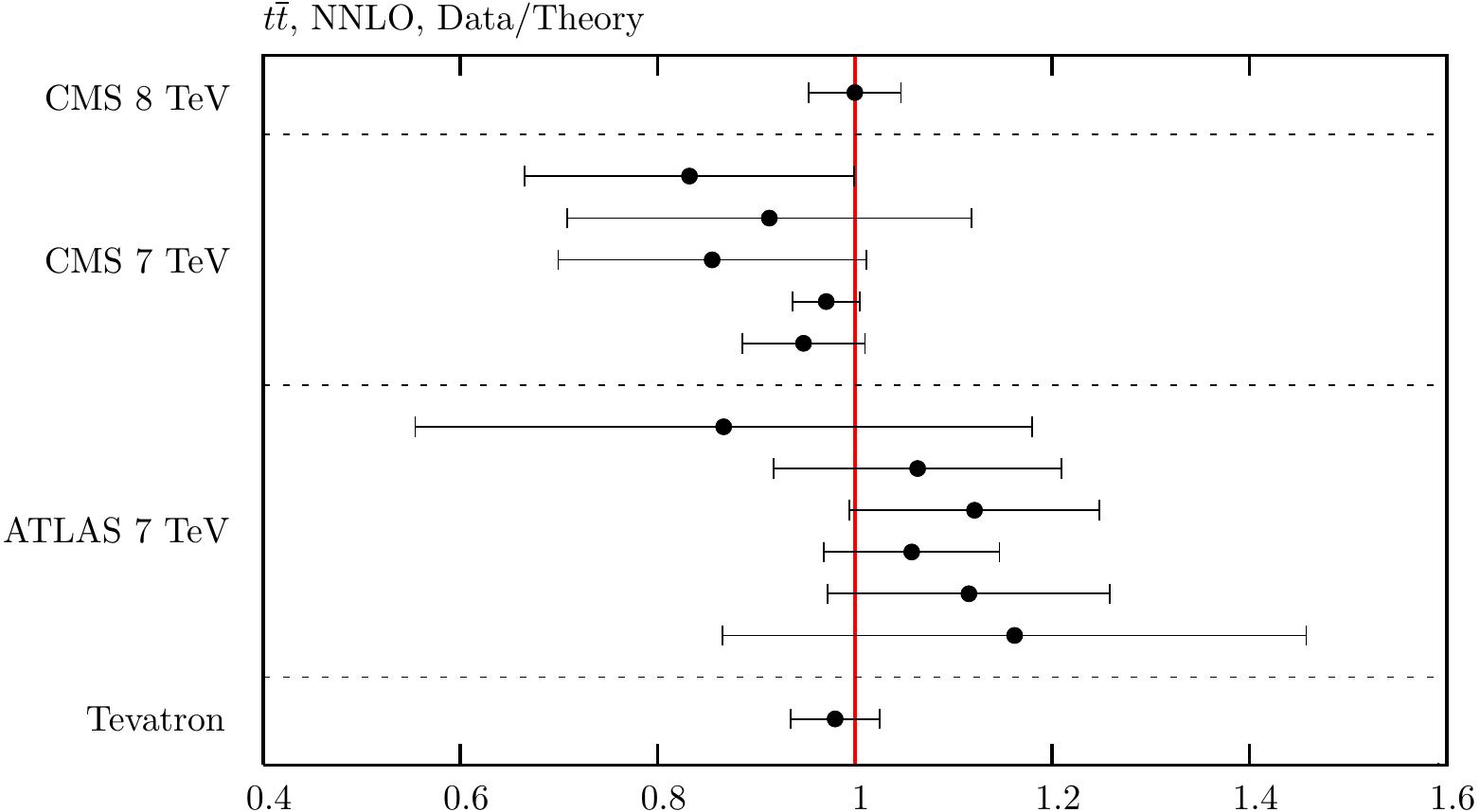}
\vspace{-0.7cm}
\caption{The MMHT fit to $\sigma_{t\bar t}$ data, figures from \cite{Harland-Lang:2014zoa}.}
\vspace{-0.2cm}
\label{Fig6}
\end{figure}

The MSTW group is renamed MMHT due
to a change in personnel. The MMHT2014 PDFs~\cite{Harland-Lang:2014zoa}
incorporate the improved parametrization and deuteron 
corrections in the MMSTWW study \cite{Martin:2012da}, and also a change in 
the heavy flavour scheme, and a change in the branching fraction 
$B_{\mu} = B(D \to \mu)$ used in the determination of the strange quark from 
$\nu N \to \mu\mu X$ data.
The updated analysis  includes new data: the combined HERA  
structure function data, improved Tevatron lepton asymmetry data, 
vector boson and inclusive jet data from the 
LHC (though LHC jet data is not included at NNLO),
and top pair cross section data from the 
Tevatron and LHC. No PDFs change dramatically in comparison to
 MSTW2008\cite{Martin:2009iq}, with the most significant changes 
being the shift in 
the small-$x$ valence quarks already observed 
in the MMSTWW study, a slight increase in the central value of the 
strange quark to help 
the fit to LHC data, and a much expanded uncertainty on the strange 
distribution. 
The PDFs are made available with 25 eigenvector pairs for $\alpha_S(m_Z^2) =0.118$ and 
0.120 at NLO and 0.118 at NNLO. However, 
$\alpha_S(m_Z^2)$ is also determined by the NLO and NNLO fits and values of 
$\alpha_S(m_Z^2)=0.1201$ and $0.1172$ respectively are found, in good agreement 
with the world average.
A dedicated study of the uncertainties in the determination of $\alpha_S(m_Z^2)$
in the MMHT2014 analysis has been presented in~\cite{Harland-Lang:2015nxa}.


\begin{wrapfigure}{r}{0.7\columnwidth}
\vspace{-0.8cm}
\centerline{\includegraphics[width=0.45\textwidth]{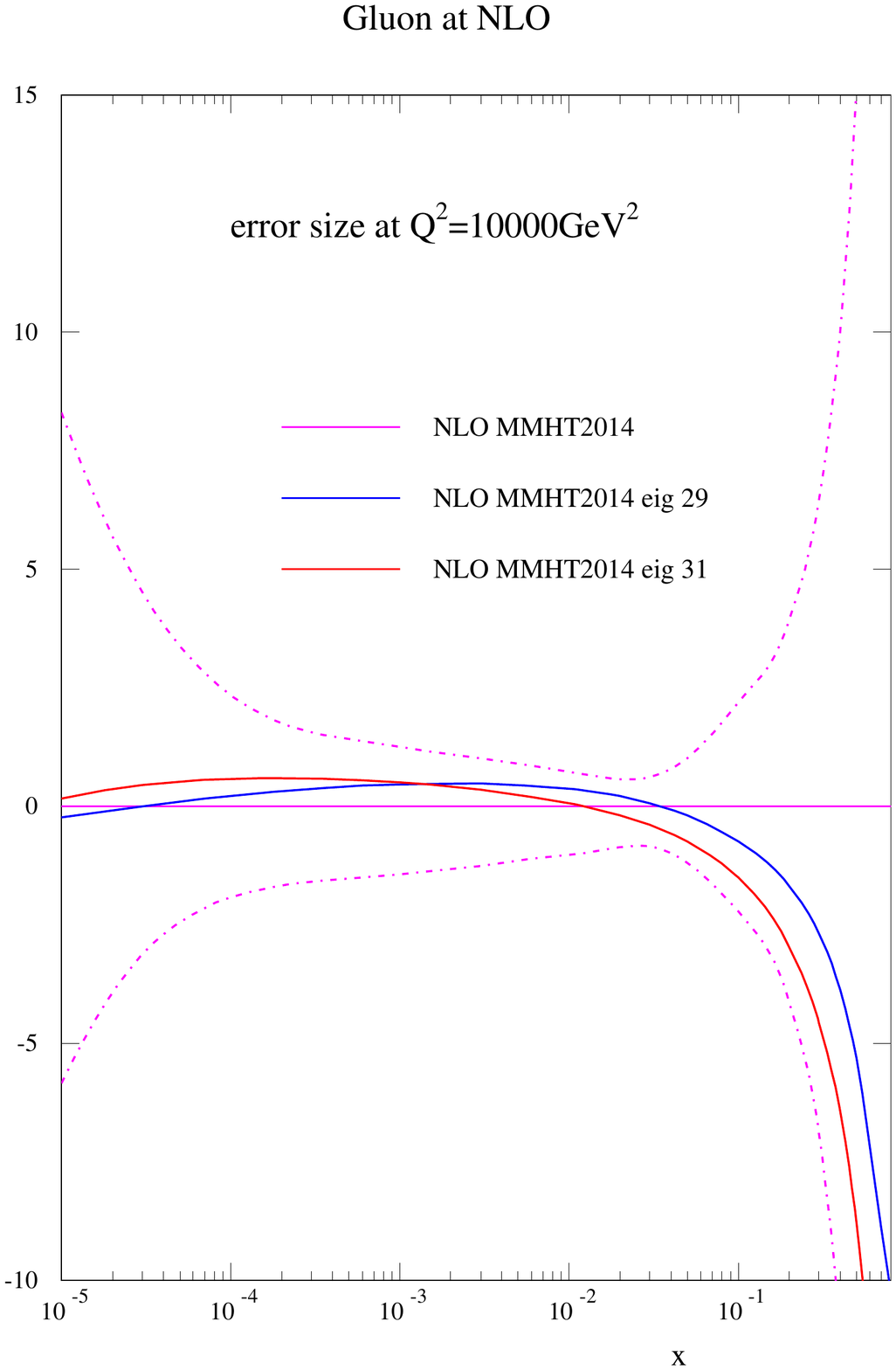}
\hspace{-0.8cm}\includegraphics[width=0.45\textwidth]{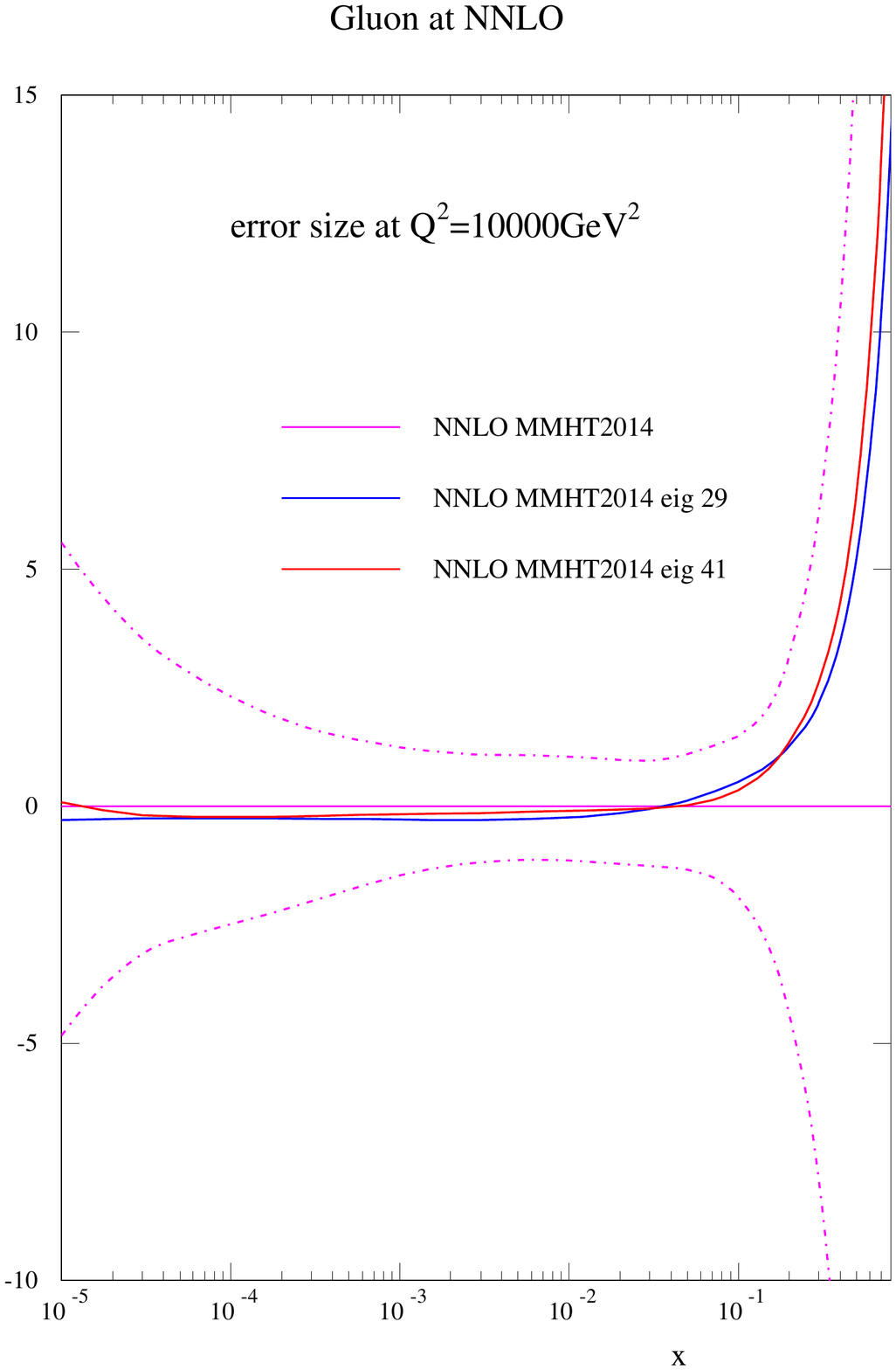}}
\vspace{-0.8cm}
\caption{Eigenvectors constraints from top cross section data for MMHT.}
\vspace{-0.2cm}
\label{Fig7}
\end{wrapfigure}

MMHT fit to data on $\sigma_{t\bar t}$ from the Tevatron
(combined cross section measurement from D0 and CDF),
and all published data from ATLAS and CMS for $7 \TeV$ and
one point at $8 \TeV$. They use $m_t = 172.5~\GeV$ with an error of 
$1~\GeV$ and with $\chi^2$
penalty applied. The predictions and the fit are good, with the
NLO fit preferring masses
slightly below $m_t = 172.5~\GeV$  and NNLO masses slightly
above, see Fig.~\ref{Fig6}.
The fit quality to $\sigma_{\bar t t}$ data alone
is very sensitive to $m_t$ and $\alpha_S(M_Z^2)$ interplay
\cite{Harland-Lang:2015nxa}.
In the NLO fit the inclusive $ t\bar t$ cross section data used
does not constrain any PDF eigenvectors. Nearly constrains 
eigenvector number 29 and 31,
both of which correspond to a decreased gluon at high $x$ only.
31 is primarily constrained by CDF jet data.
In the NNLO fit the inclusive $ t\bar t$ cross section
constrains one eigenvector, number 29 and (nearly) 41.
Both correspond to increased gluon at high $x$ only.
The eigenvectors are shown in Fig.~\ref{Fig7}

\begin{figure}[]
\vspace{-0.2cm}
\centerline{\includegraphics[width=0.46\textwidth]{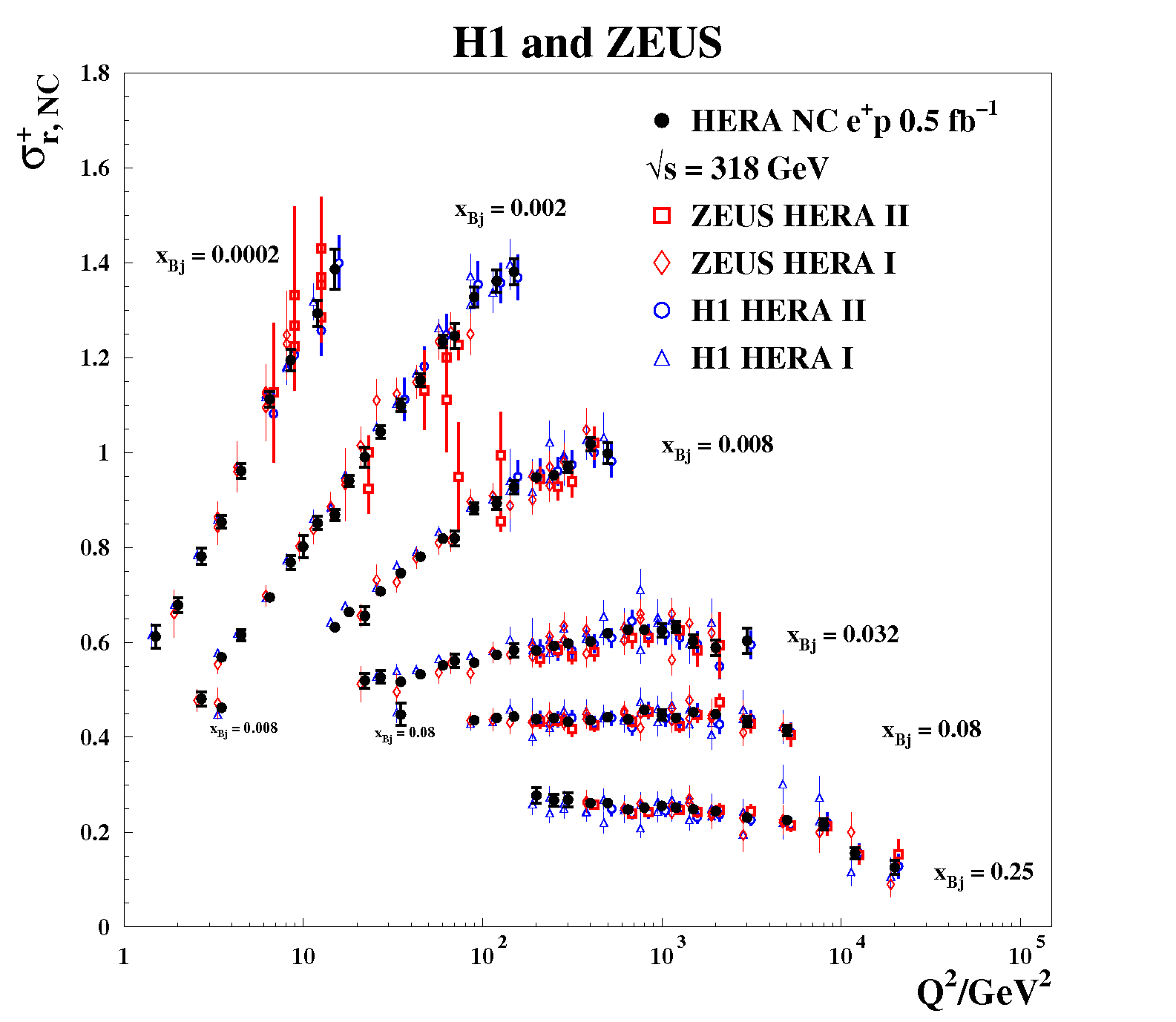}
\includegraphics[width=0.52\textwidth]{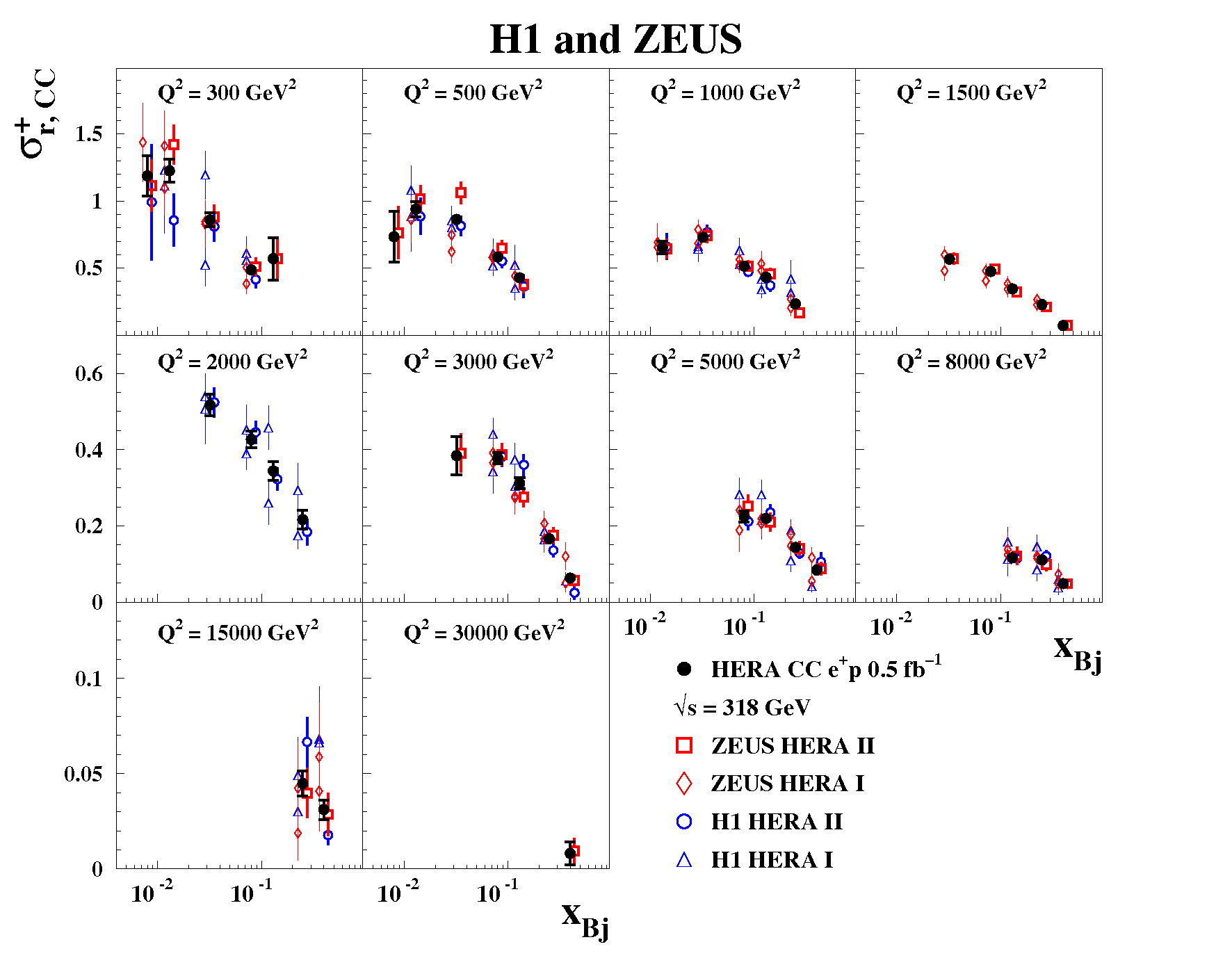}}
\vspace{-0.1cm}
\caption{Neutral(left) and charged (right) current data 
from the final HERA combination, figure from \cite{Abramowicz:2015mha}.}
\vspace{-0.2cm}
\label{Fig8}
\end{figure}

The data for $t \bar t$ differential
distributions are not currently used in PDF determinations 
as they did not meet cut-off dates for data inclusion and also
had missing NNLO corrections which may be important.
In comparison with existing PDFs at NLO the $y_{\bar t t}$ distribution tends 
to be very good, but the $p_{t}$ distribution off in shape, while $m_{\bar t t}$
is somewhere in between). It is interesting to see the NNLO corrections
\cite{Czakon:2015owf} improve the comparison to the $p_T$ distribution 
markedly.

Since these updates a  
HERA combination of all inclusive structure function measurements
from Runs I and II has been presented \cite{Abramowicz:2015mha}, and 
included in the HERAPDF2.0 set. The improved data can be seen in 
Fig.~\ref{Fig8}.
The resulting HERAPDF set has considerably reduced uncertainties, and 
a much improved constraint on flavour decomposition
at moderate and high $x$ due to the difference between neutral current 
$e^+$ and $e^-$ 
cross sections, and to much more precise charged current data. 
The running at different energies gives sensitivity to 
$F_L(x,Q^2)$ which constrains the gluon.

\begin{figure}
\vspace{-0.1cm}
\includegraphics[width=0.5\textwidth]{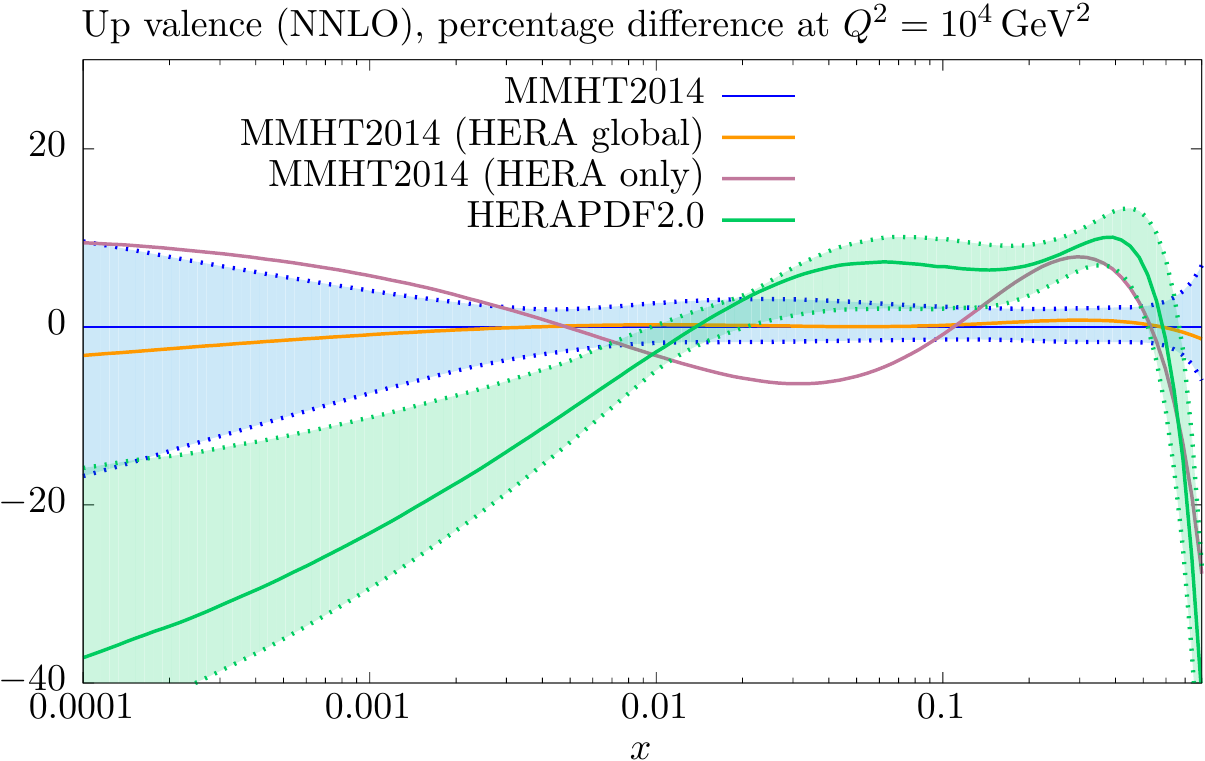}
\includegraphics[width=0.5\textwidth]{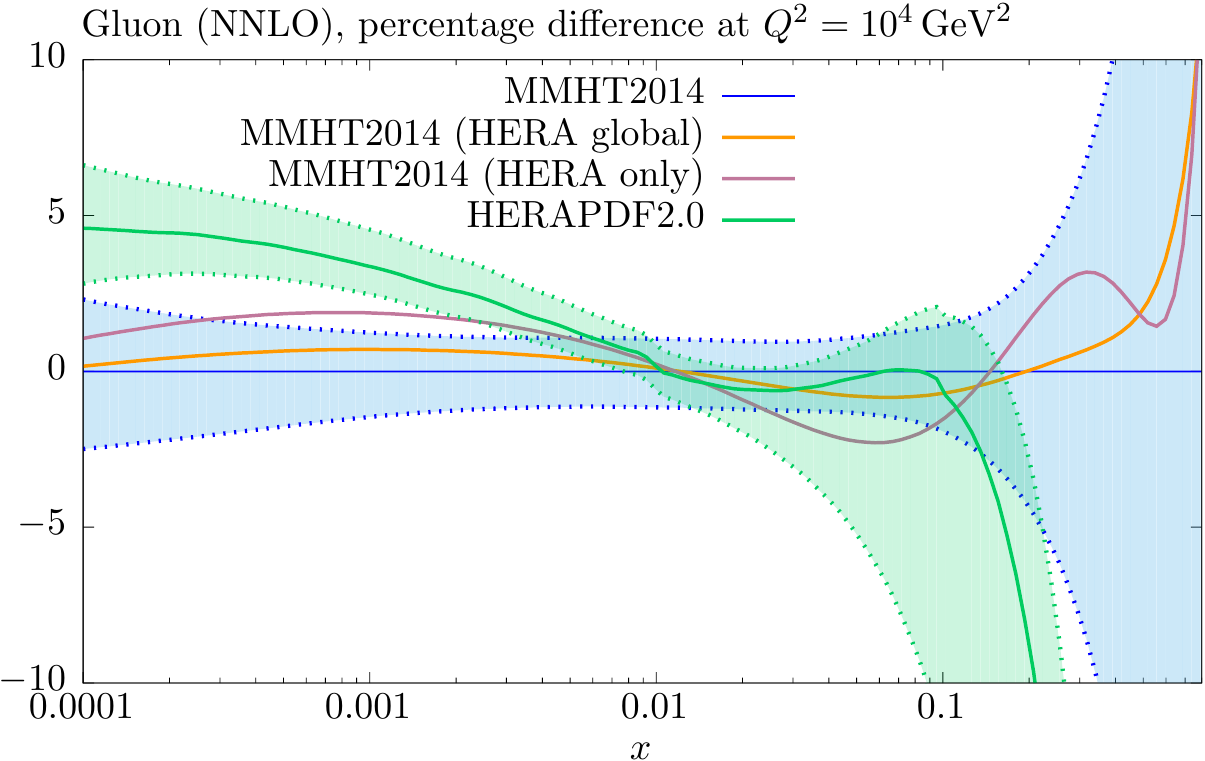}\\
\includegraphics[width=0.485\textwidth]{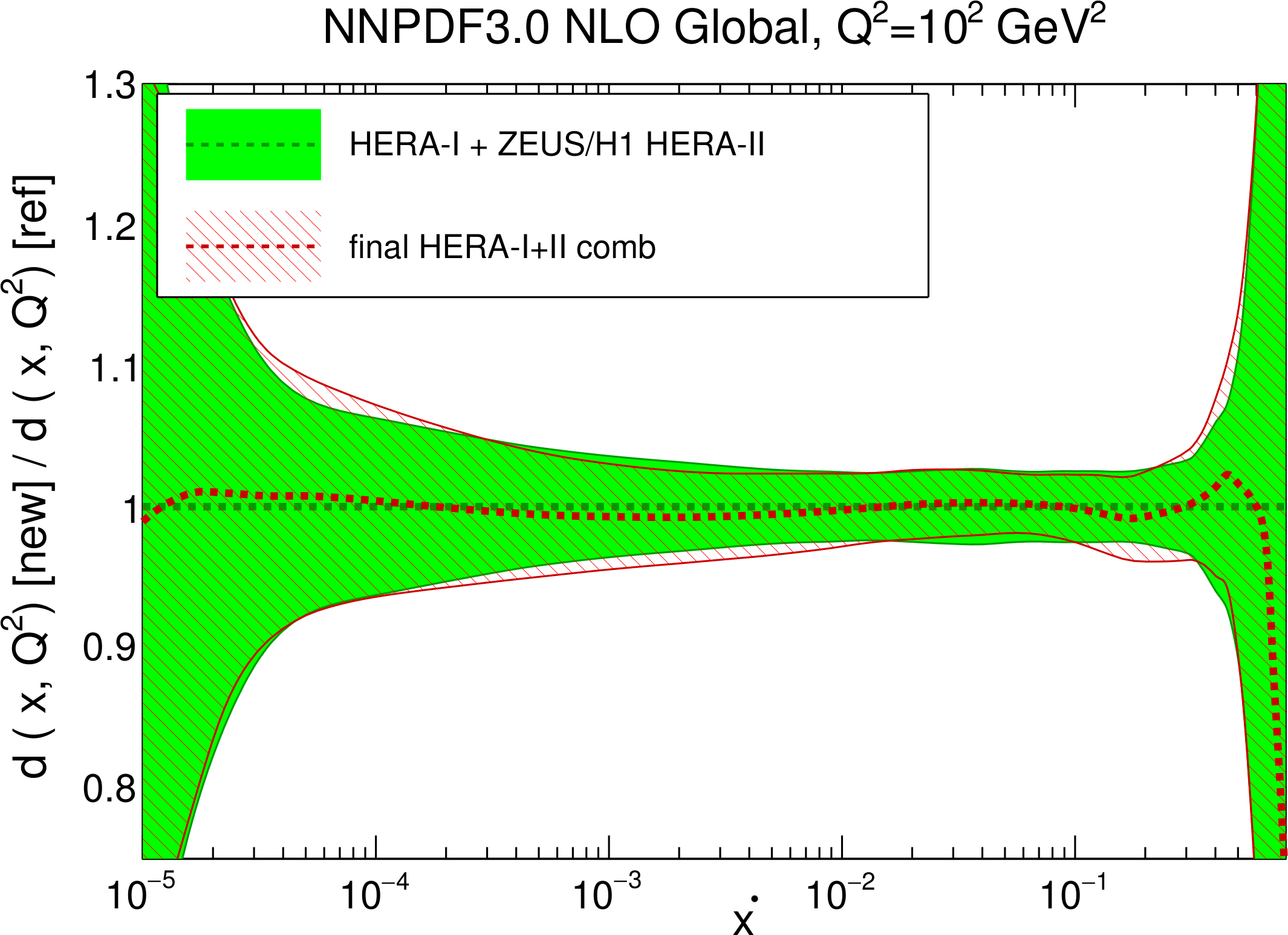}
\hspace{0.3cm}\includegraphics[width=0.485\textwidth]{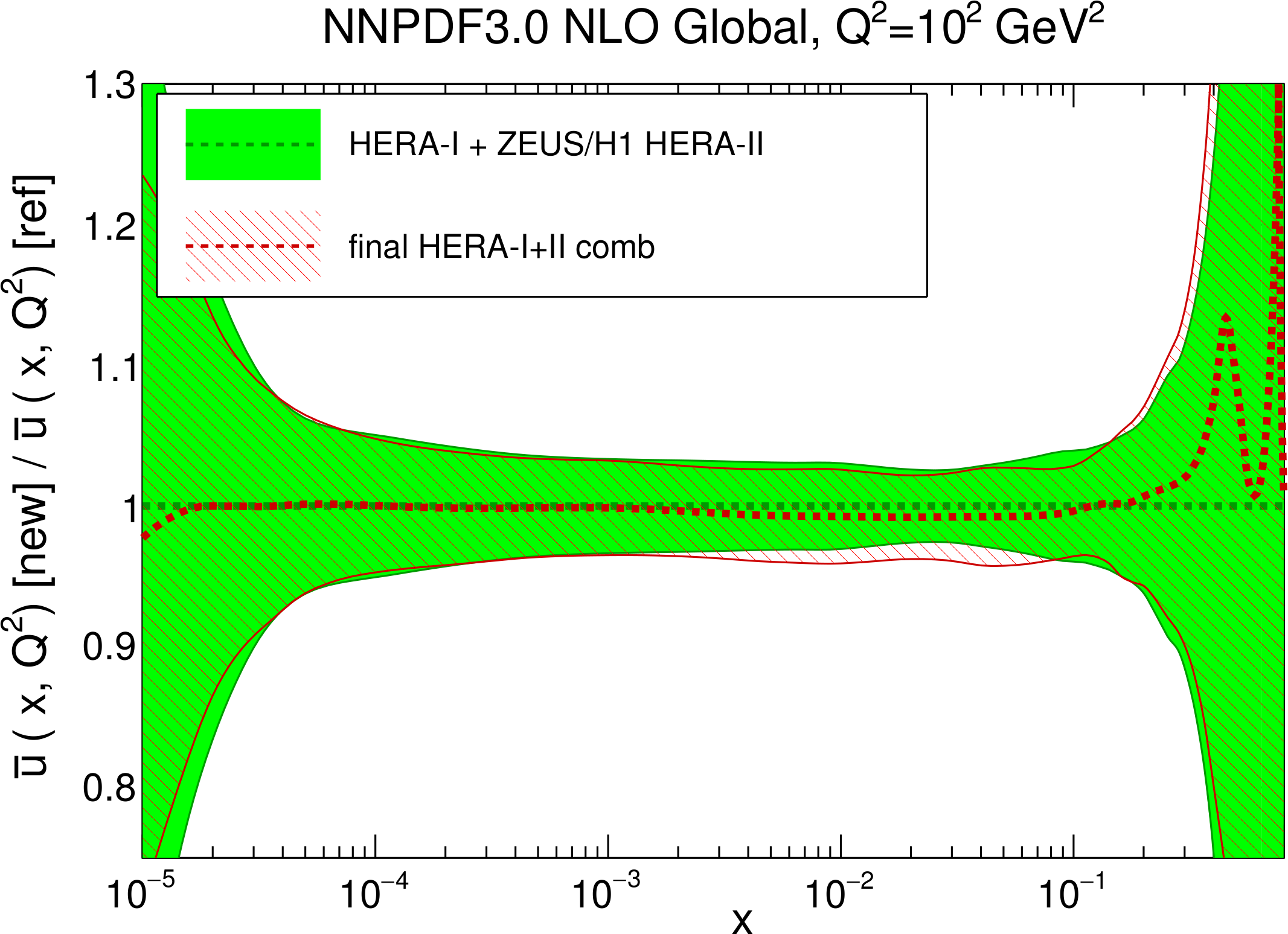}
\vspace{-0.65cm}
\caption{MMHT (top) (figures from \cite{Thorne:2015caa}) and NNPDF (bottom) 
(figures from \cite{Rojo:2015nxa}) PDFs with the inclusion of the final HERA combined data.}
\vspace{-0.2cm}
\label{Fig9}
\end{figure}

These HERA combined data have now been included in global fits.
Good fits, with little deterioration for other data are obtained
for both MMHT \cite{Thorne:2015caa,Harland-Lang:2016yfn} and NNPDF \cite{Rojo:2015nxa}. 
These also result in small changes 
in the central PDFs and 
uncertainties, as shown in Fig.~\ref{Fig9}. Hence, there is no imperative 
to provide immediate further updates since these will appear soon due to new 
LHC data.

\begin{figure}
\vspace{-0.2cm}
\centerline{\includegraphics[width=0.49\textwidth]{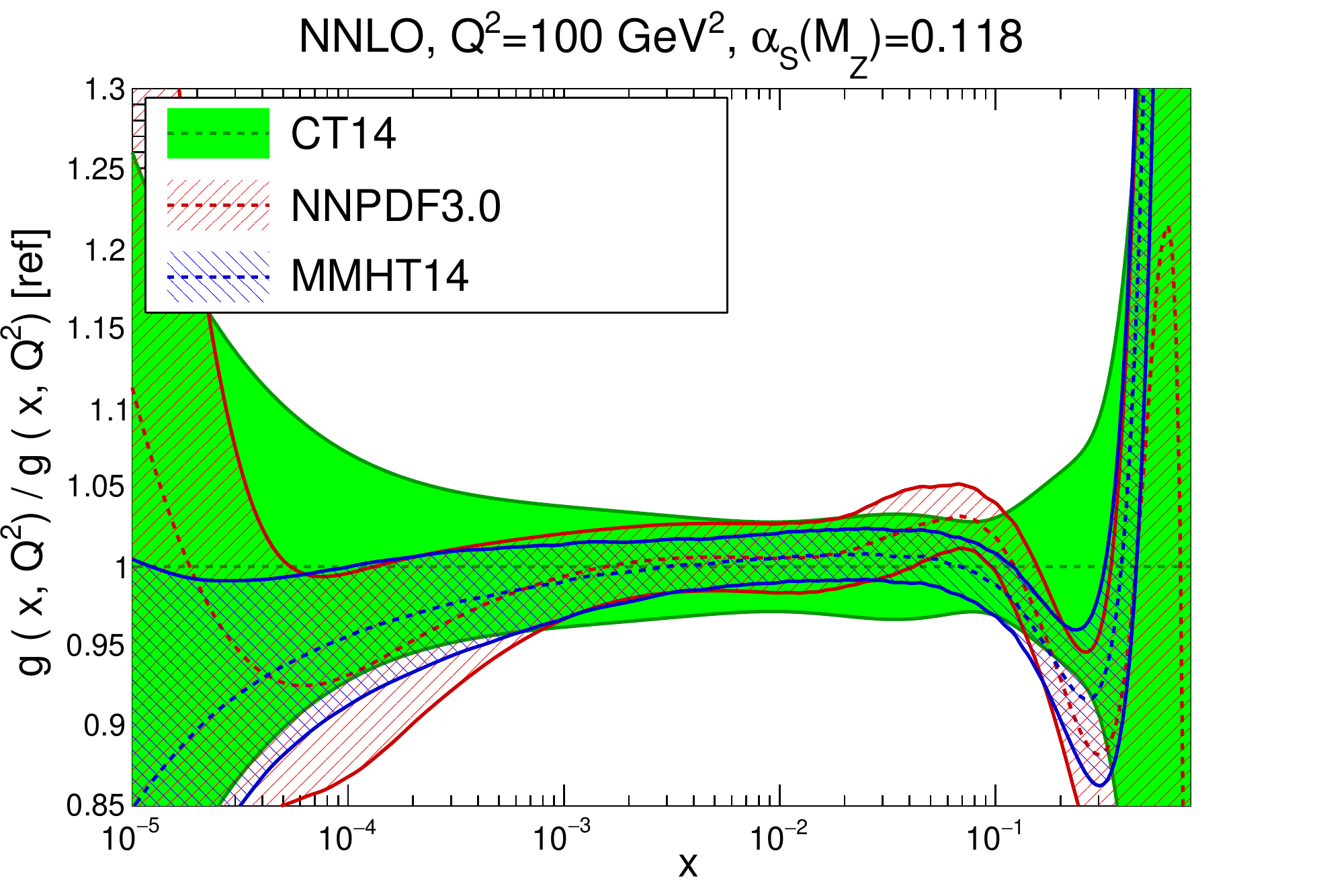}
\includegraphics[width=0.49\textwidth]{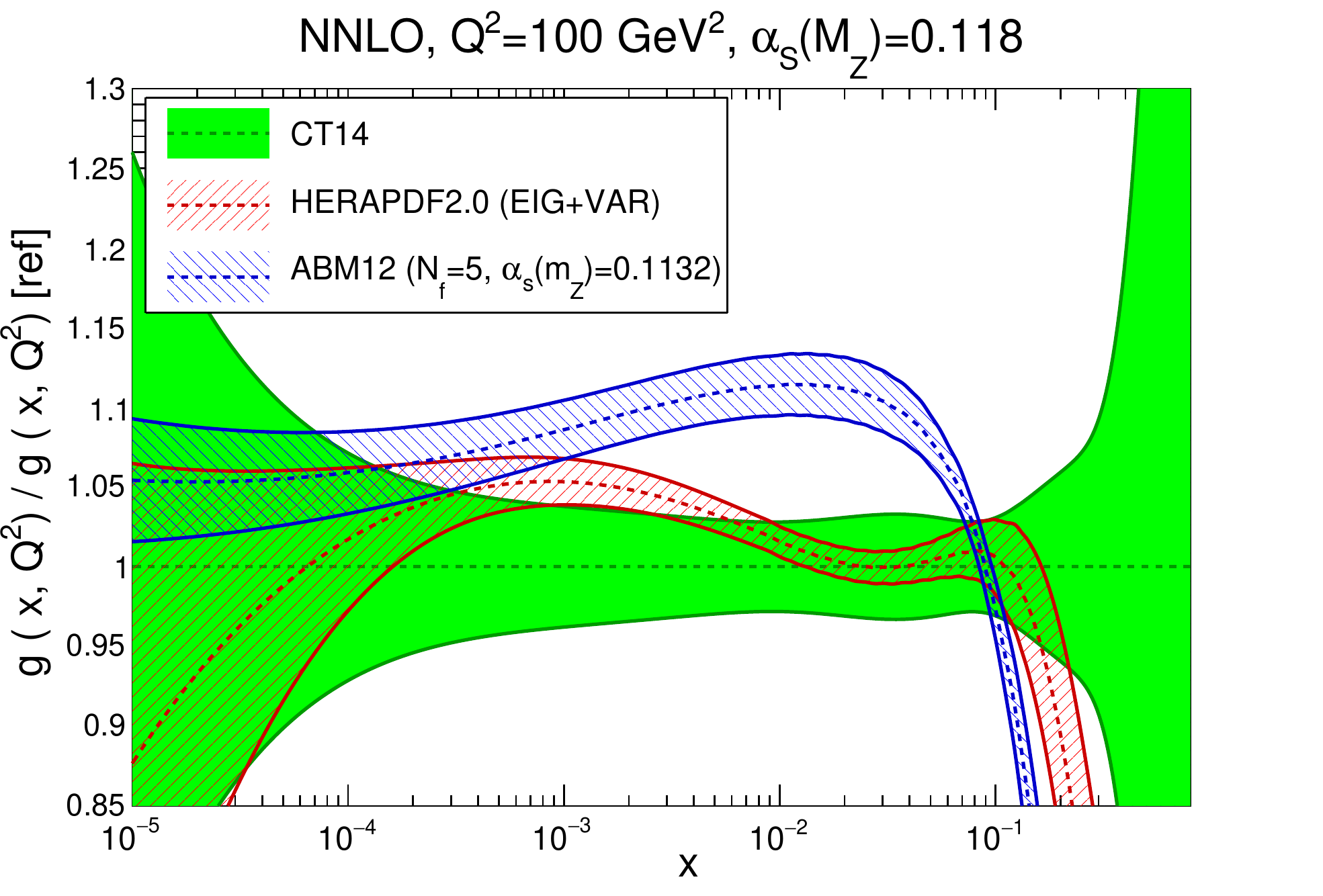}}
\vspace{-0.2cm}
\centerline{\includegraphics[width=0.49\textwidth]{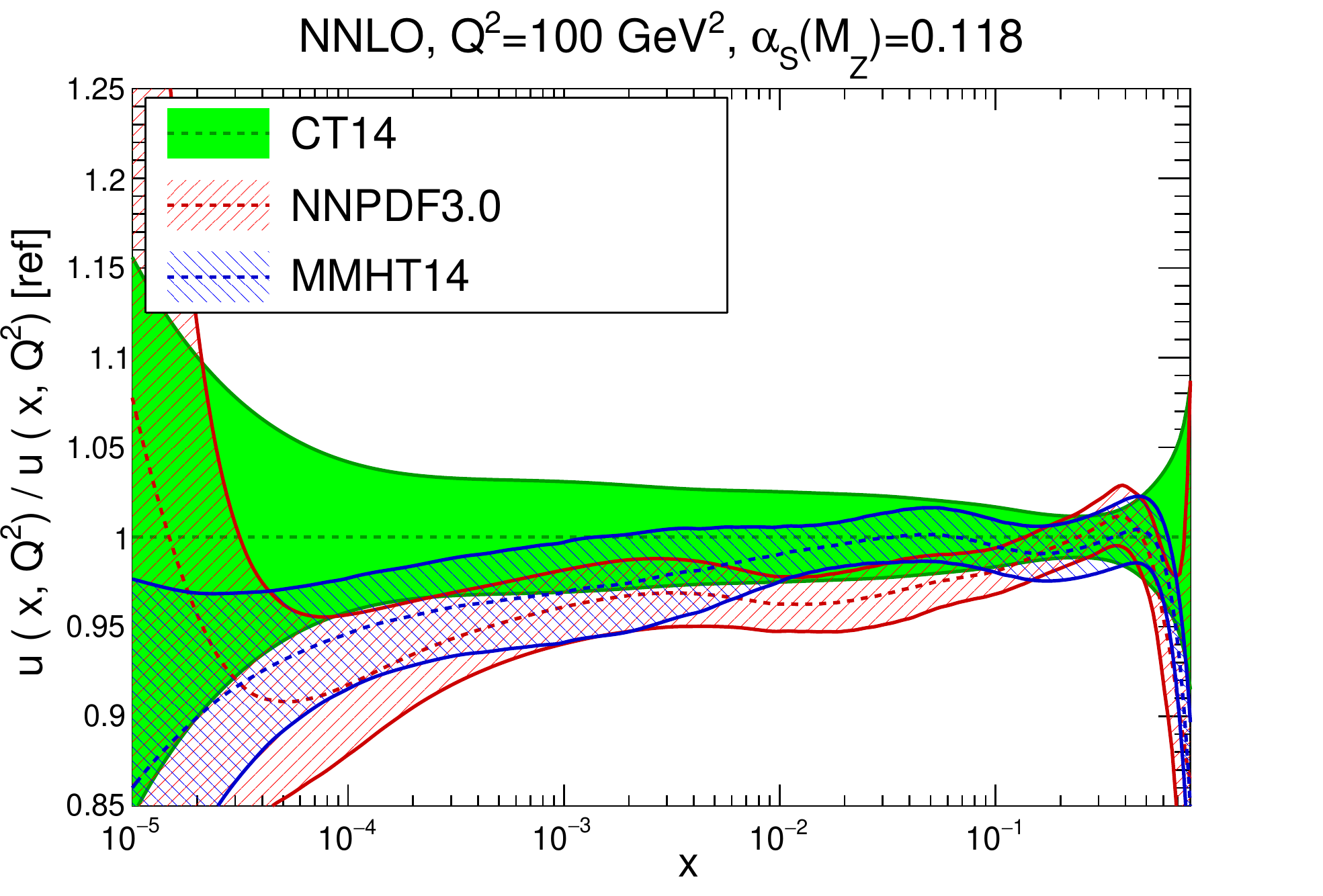}
\includegraphics[width=0.49\textwidth]{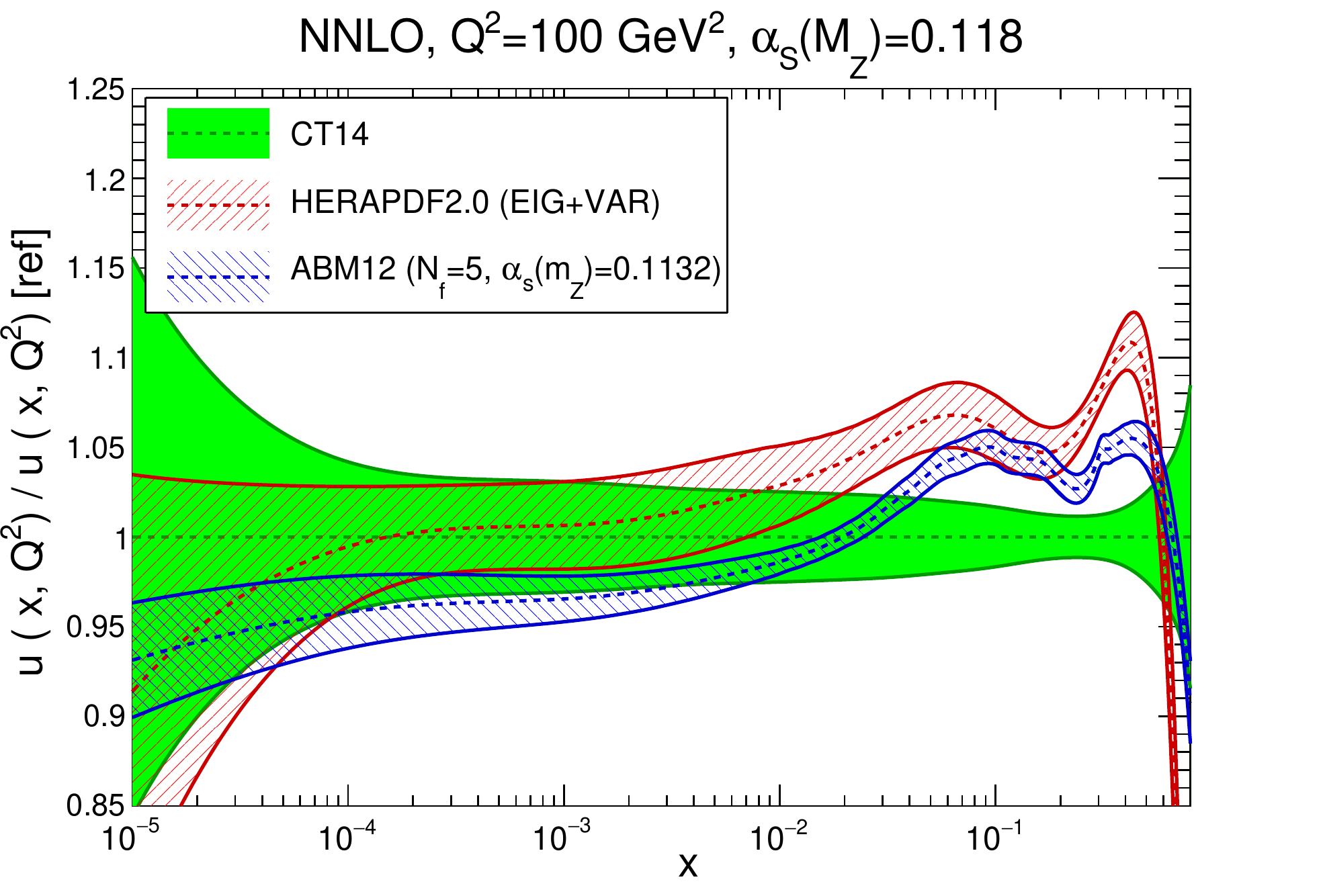}}
\vspace{-0.2cm}
\centerline{\includegraphics[width=0.49\textwidth]{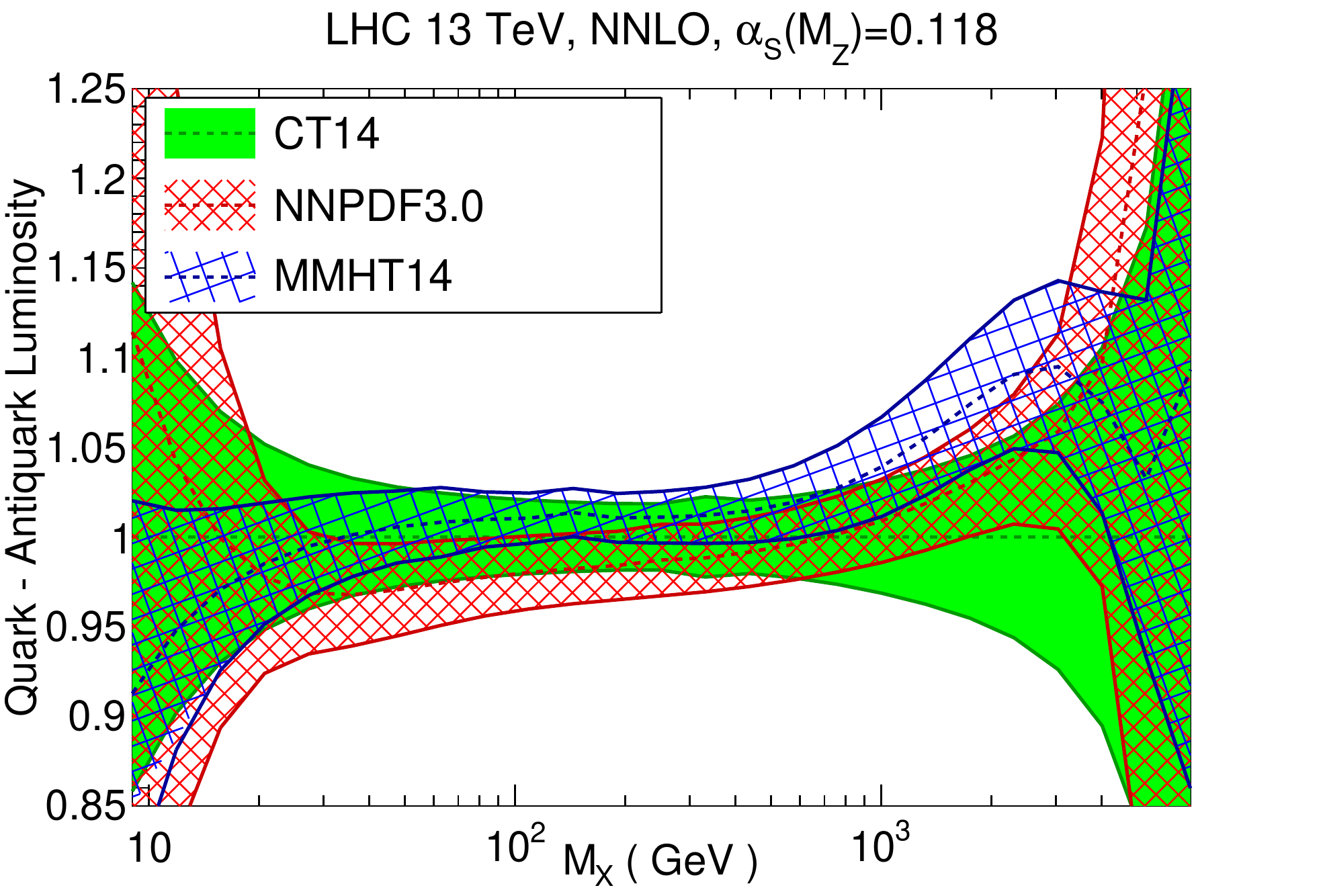}
\includegraphics[width=0.49\textwidth]{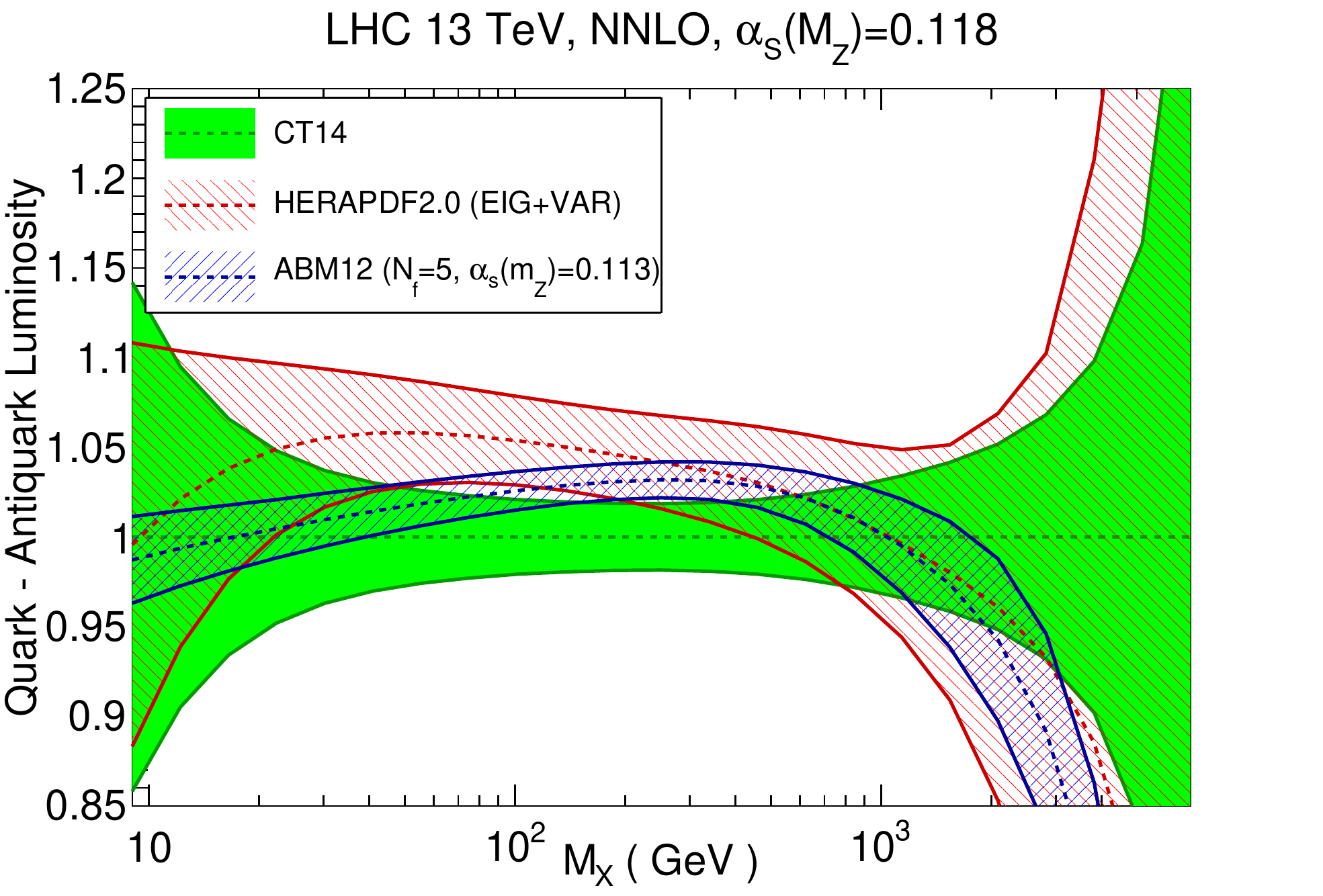}}
\vspace{-0.2cm}
\centerline{\includegraphics[width=0.49\textwidth]{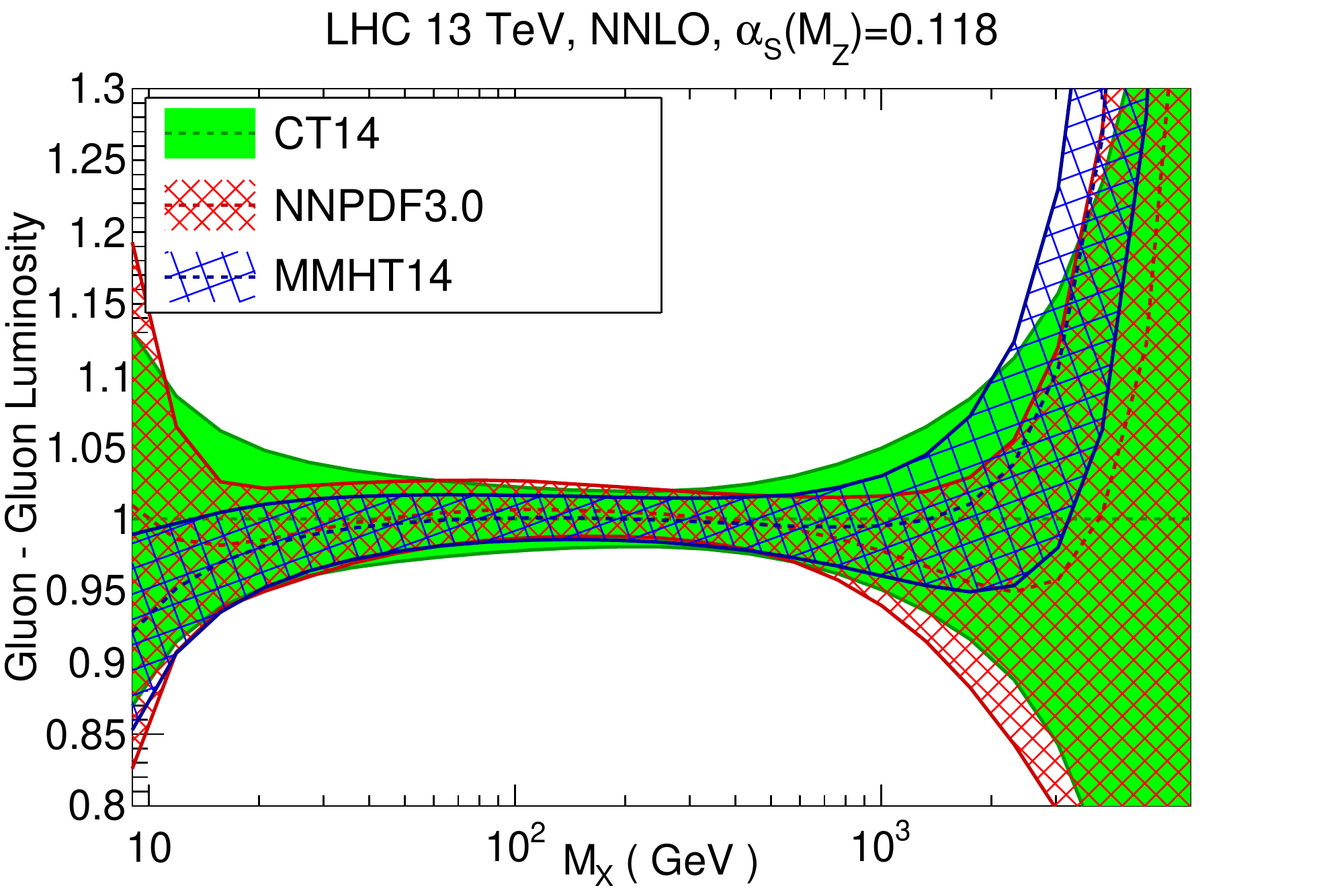}
\includegraphics[width=0.49\textwidth]{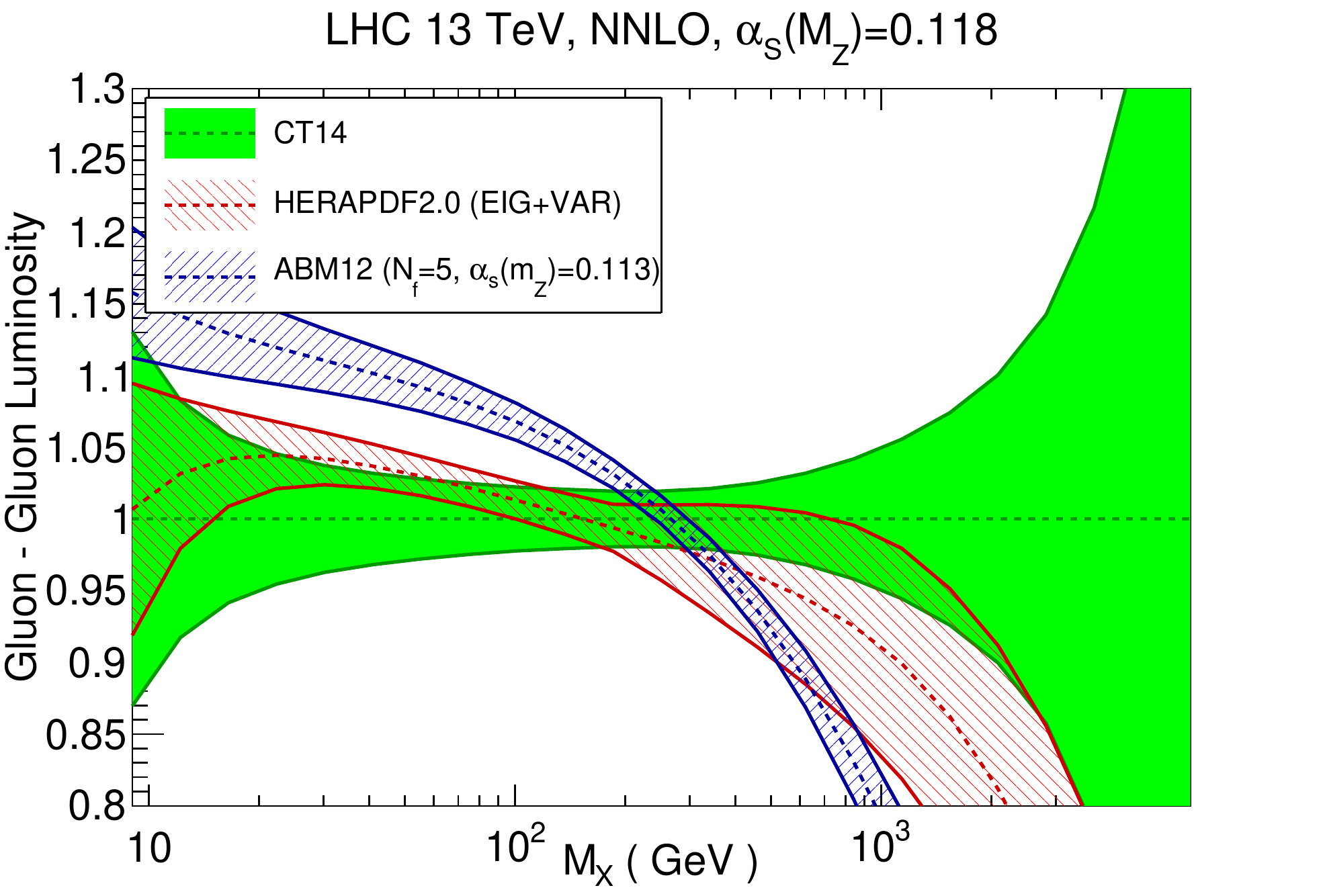}}
\vspace{-0.3cm}	 
\caption{The comparison of different PDFs (top two plots) and parton luminosities (lower two plots), figures from \cite{Butterworth:2015oua}.}
\vspace{-0.9cm}	 
\label{Fig10}
\end{figure}

The comparison between the most recent versions of the different PDF sets
is shown for the gluon and up quark in the upper of Fig.~\ref{Fig10}. 
There is now excellent agreement between CT14, 
MMHT2014 and NNPDF3.0, much better than in the previous versions of these PDF sets, but there is still some significant differences in central
values and uncertainty between the other PDF sets.
The comparison of PDF luminosities is shown also shown in the lower of 
Fig.~\ref{Fig10}. 
The $gg$ luminosity now in almost perfect agreement for the three ``global''
sets, but some variation is seen in quark (antiquark) luminosities.

\section{Combination of PDF sets}

\begin{figure}
\vspace{-0.1cm}
\centerline{\includegraphics[width=0.9\textwidth]{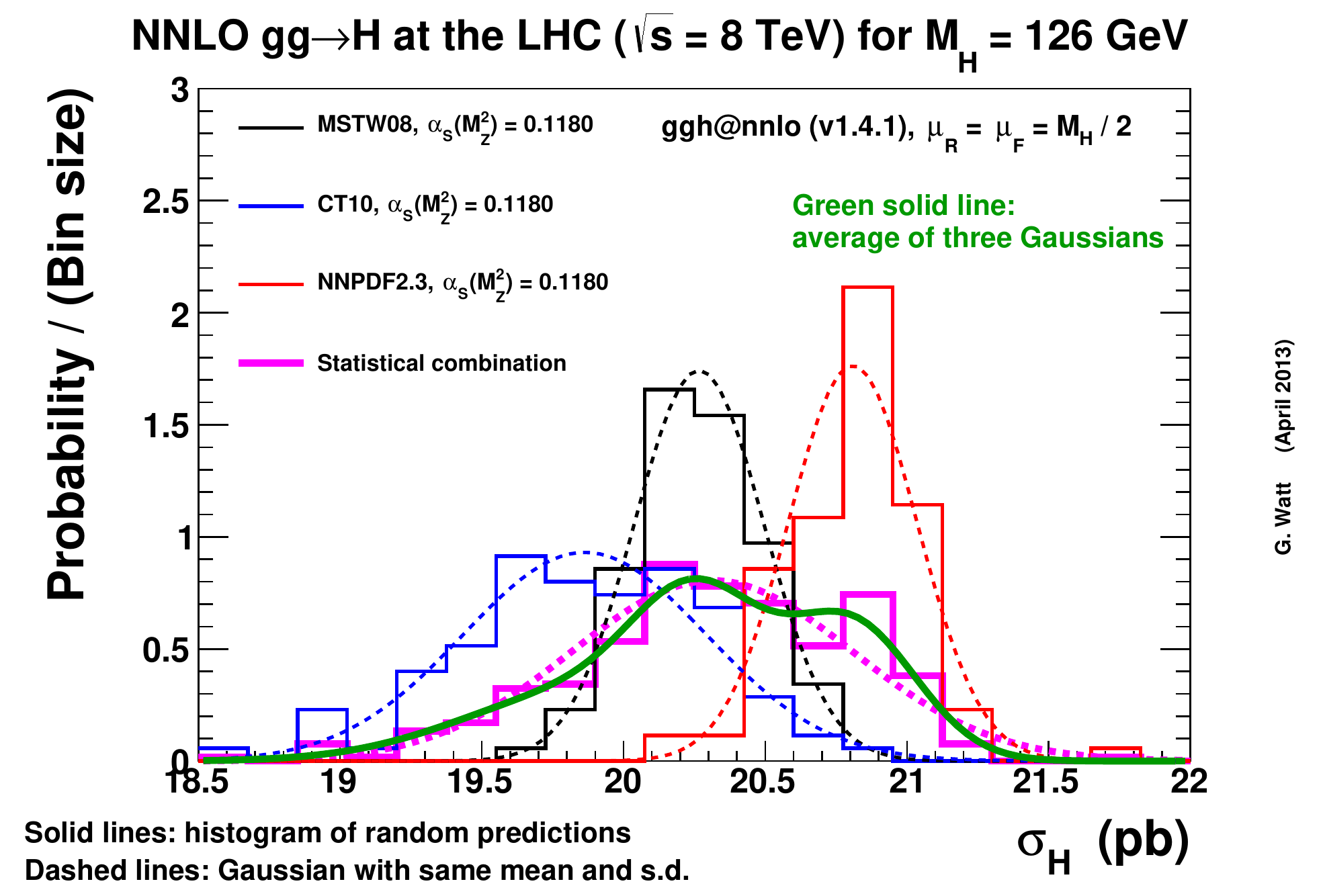}}
\vspace{-0.1cm}
\caption{Combination of distributions for $\sigma_{gg \to H}$
(plot by G. Watt. \cite{WattPDF4LHC}).}
\vspace{-0.3cm}
\label{Fig13}
\end{figure}

It is not obvious how to combine different ``Hessian'' PDF sets.
However, it is now known how to generate ``random'' PDF sets directly from 
the representation in terms of eigenvectors \cite{Watt:2012tq}
\begin{equation}
F(\mathcal{S}_k) = F(S_0) + \sum_{j}\left[\!F(S_j^\pm)- F(S_0)\!\right] 
|R_{jk}| \nonumber
\end{equation}
Hence, one  can combine different PDF sets either at PDF level or predictions. 
The latter is shown using the last round of global PDFs for the Higgs cross 
section in Fig.~\ref{Fig13}, and can be applied to the PDFs at a particular 
$x$ and $Q^2$ value in the same manner.

\begin{figure}
\vspace{-0.2cm}
\centerline{\includegraphics[width=0.5\textwidth]{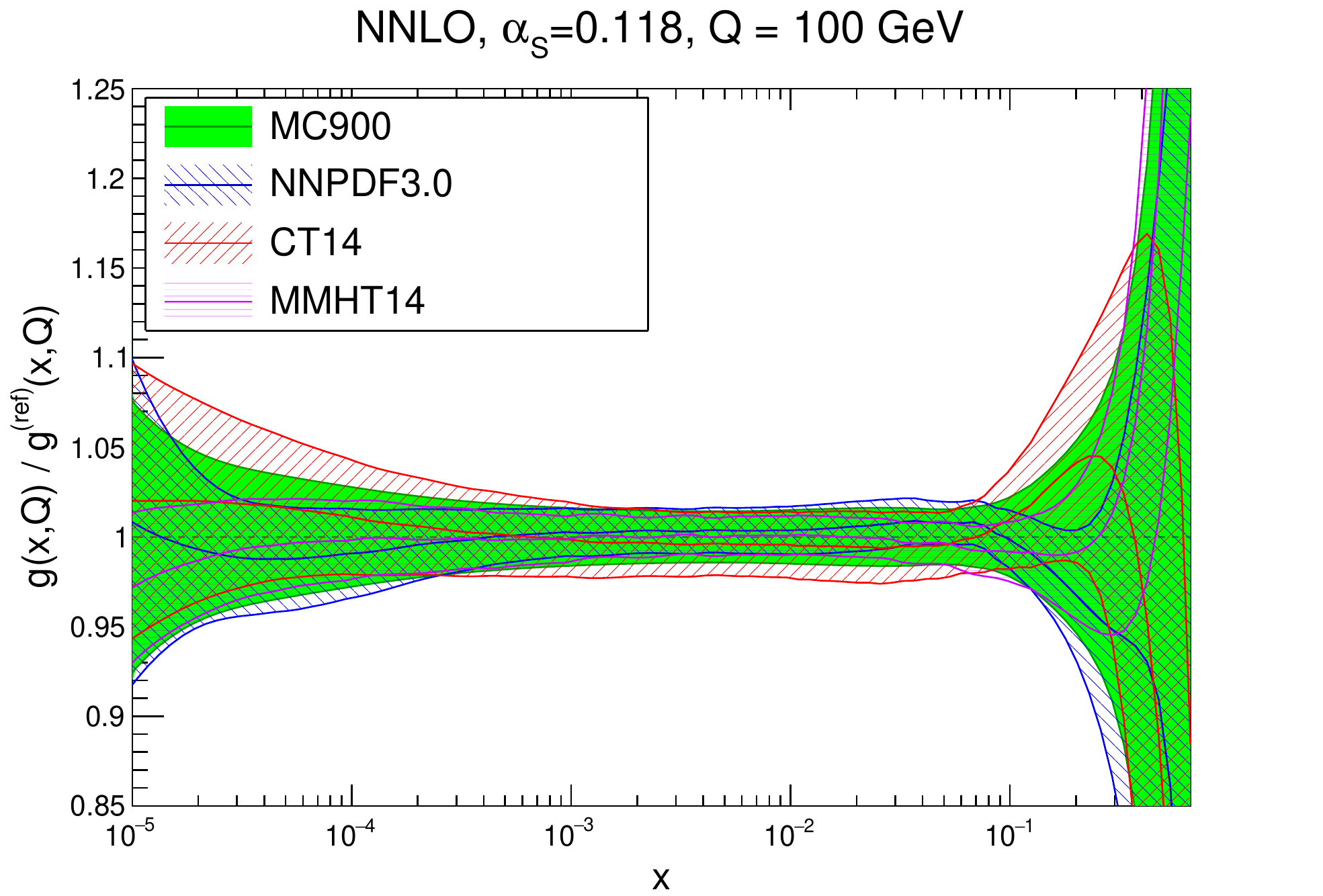}
\includegraphics[width=0.5\textwidth]{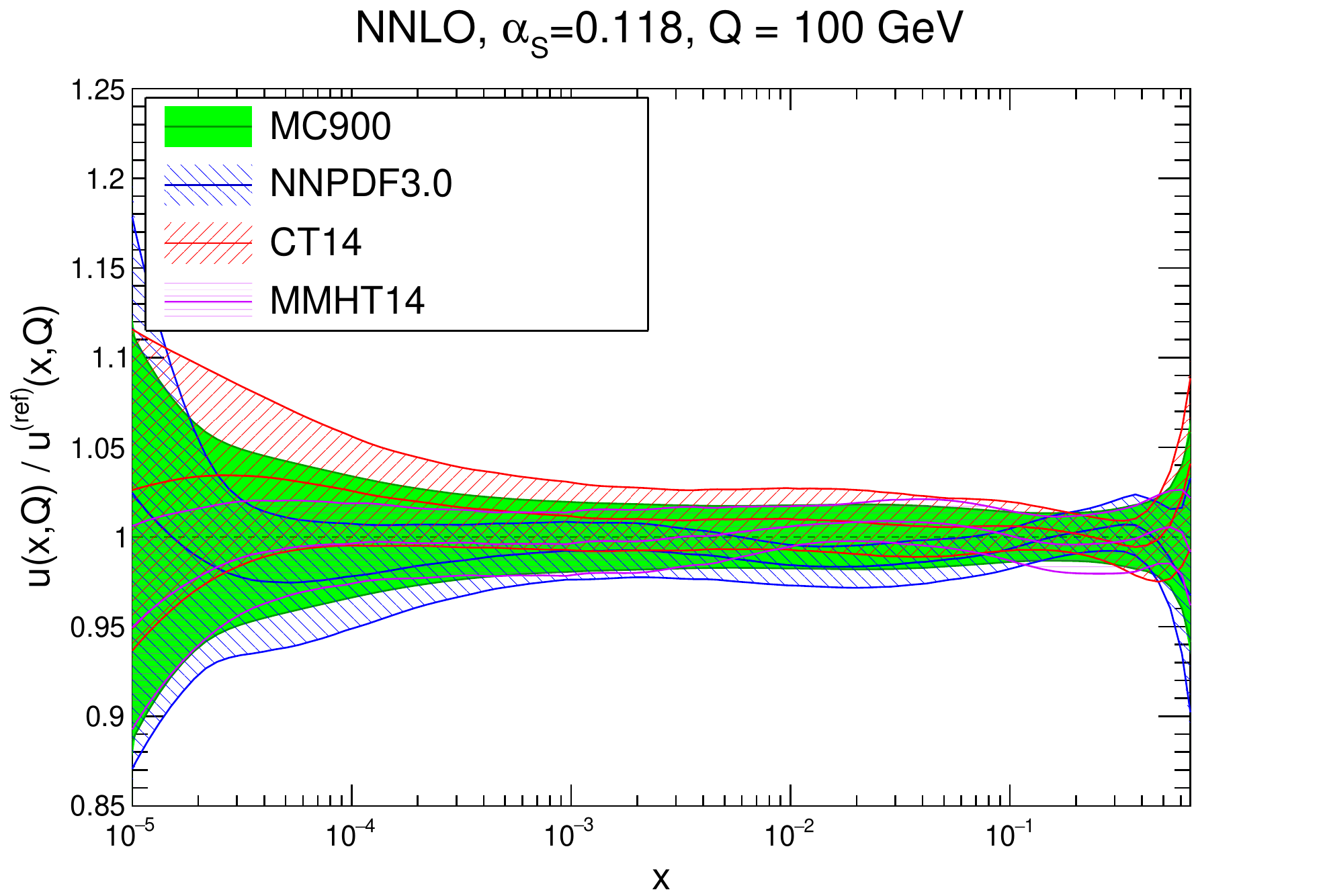}}
\vspace{-0.2cm}
\centerline{\includegraphics[width=0.5\textwidth]{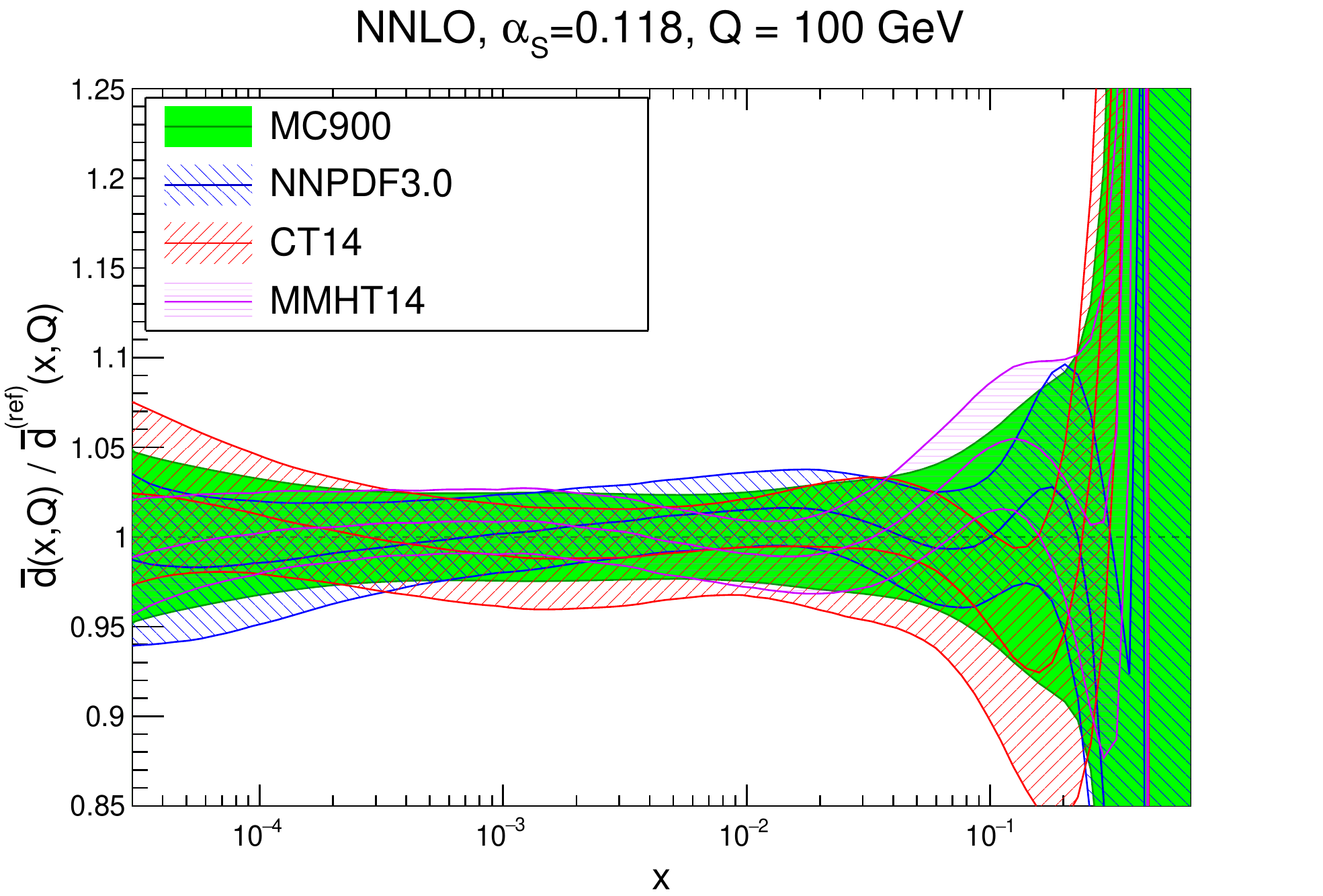}
\includegraphics[width=0.5\textwidth]{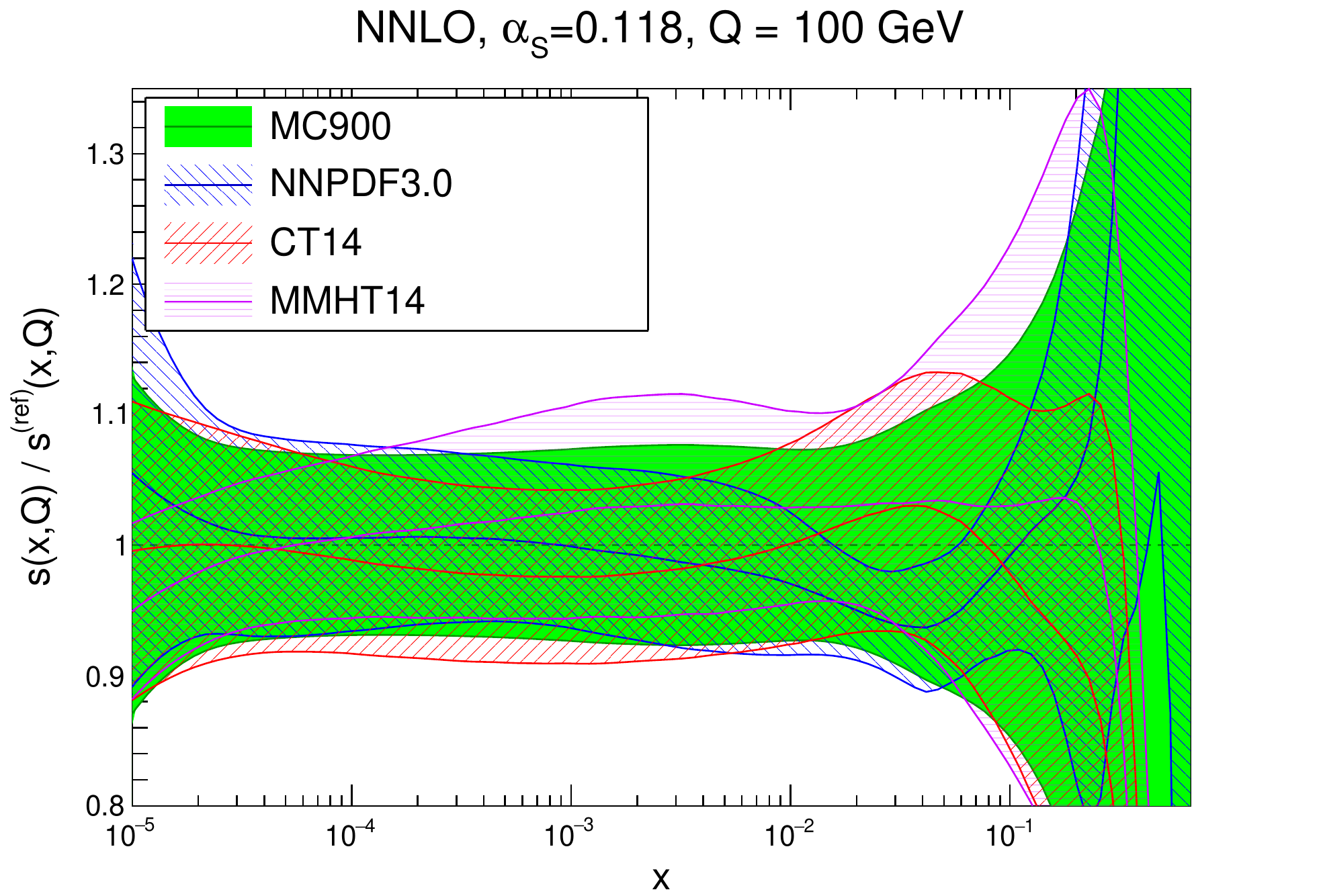}}
\vspace{-0.4cm}
\caption{The combination of 300 randomly distributed sets of each 
of the CT14, MMHT2014 and NNPDF3.0 PDF sets, figures from \cite{Butterworth:2015oua}.}
\vspace{-0.1cm}
\label{Fig14}
\end{figure}

The application to the combination of the CT14, MMHT2014 and NNPDF3.0 PDFs
is shown in Fig.~\ref{Fig14}. It works well 
if the PDFs are fairly compatible - both in central value
and uncertainty - giving the mean of the central values and a 
spread which combines the individual PDF uncertainties and the variation 
in the PDFs. Following this initial development the 
{Meta-PDF} approach \cite{Gao:2013bia} subsequently 
showed how refit the combination in terms of a large number Monte Carlo PDFs
to a functional form, and hence convert the combination to Hessian set with 
a relatively small number of eigenvector sets. Further developments showed 
how to compress the Monte Carlo set to a smaller number 
\cite{Carrazza:2015hva} and how to use 
the Monte Carlo sets in the combination as a basis for an extremely precise 
Hessian representation (MC-H) \cite{Carrazza:2015aoa}.

\section{The New PDF4LHC Prescription}

\begin{figure}
\vspace{-0.0cm}
\centerline{\includegraphics[width=0.5\textwidth]{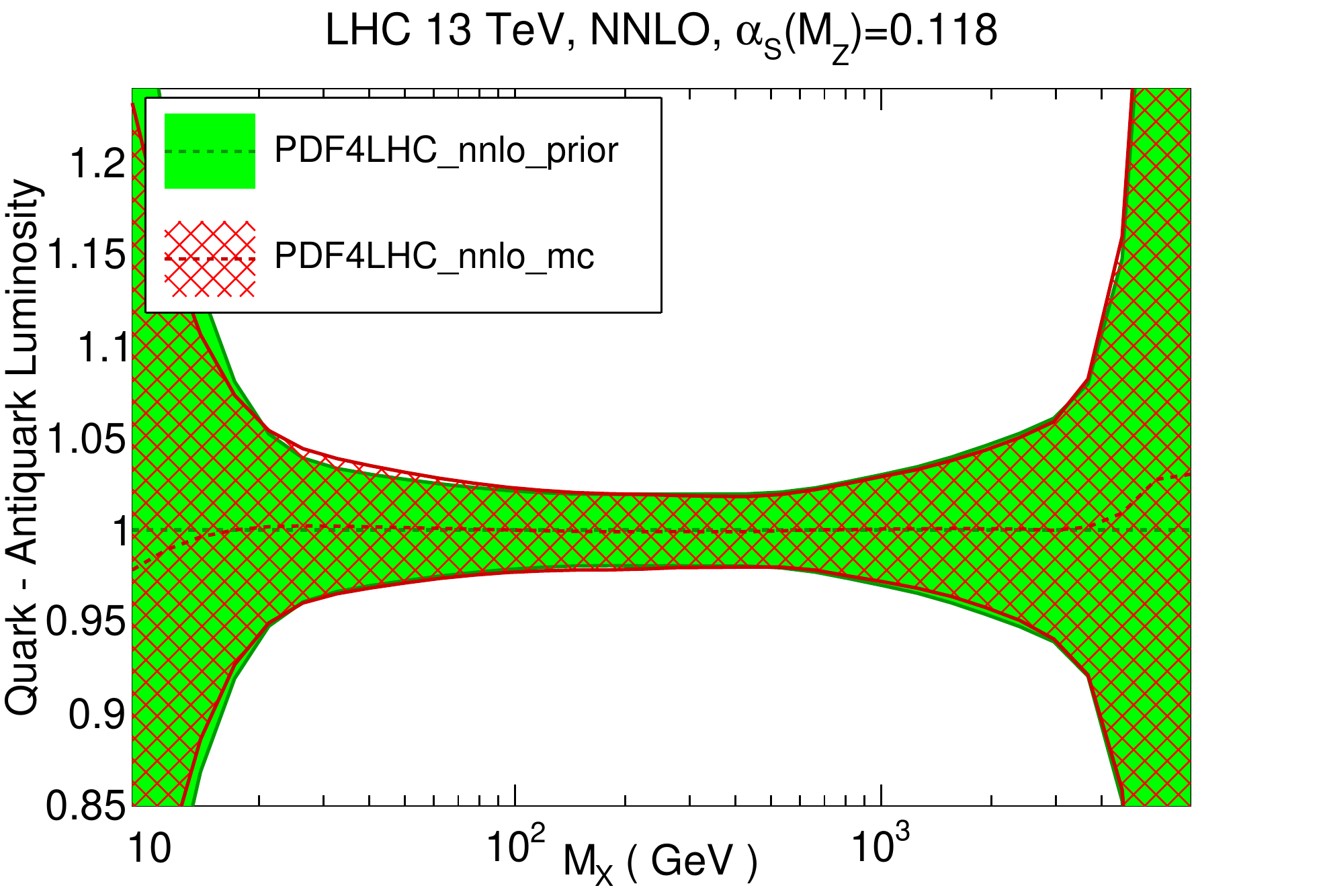}
\includegraphics[width=0.5\textwidth]{qq_mc900_vs_cmc100}}
\centerline{\includegraphics[width=0.5\textwidth]{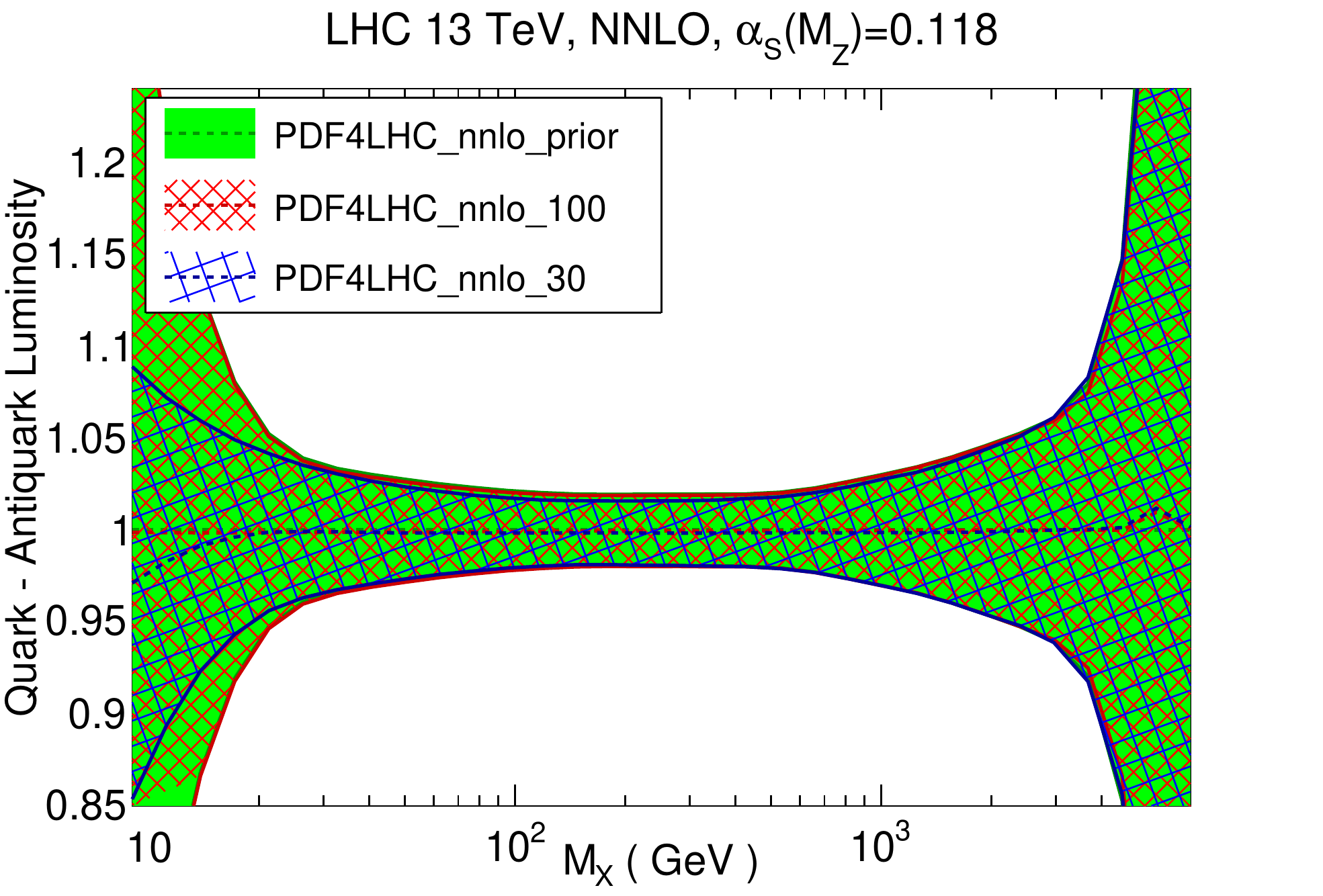}
\includegraphics[width=0.5\textwidth]{qq_mc900_vs_mch100_vs_meta30}}
\vspace{-0.1cm}
\caption{Comparison of PDF luminosities for Monte Carlo compression (top) 
and Hessian compression (bottom), figures from \cite{Butterworth:2015oua}.}
\vspace{-0.2cm}
\label{Fig15}
\end{figure}

The improved agreement of the global PDF sets and the means of combining 
them in a more statistically robust fashion allows for an update in the 
previous PDF4LHC prescription \cite{Botje:2011sn} for combining PDFs when 
a single prediction representing a 
reasonable average prediction and quite conservative uncertainty is required. 
The sets entering into the combination must satisfy requirements, i.e.
be compatible for combination, and at present CT14, MMHT2014 and NNPDF3.0 are 
included. It has been agreed that this should be for the 
common value of the coupling $\alpha_S(M_Z^2)=0.118$.
The recommendation now allows the use of a single combined PDF set in 
either Monte Carlo
or Hessian form \cite{Butterworth:2015oua}: 
Monte Carlo - A set of PDF replicas is delivered, where
the mean is the central value and the standard deviation the uncertainty;
Hessian - a central set and eigenvectors representing
orthogonal sources of uncertainty are delivered, and the uncertainty 
obtained by summing each uncertainty source in quadrature.
In each case a single combined set at both $\alpha_S(M_Z^2)=0.1165$ 
and $\alpha_S(M_Z^2)=0.1195$ is provided to
give the $\alpha_S(M_Z^2)$ uncertainty
(i.e. $\Delta \alpha_S(M_Z^2)=0.0015$) to be added in quadrature
with other uncertainties.

Three different options are provided along with suggestions for when 
they should be used:

\noindent {\bf PDF4LHC15-mc:} A compressed {\bf Monte Carlo} set with
$N_{\rm rep}=100$ \cite{Carrazza:2015hva}. Contains non-gaussian features --
important for searches at high masses (high $x$). See Fig. \ref{Fig15} for 
the compressed set compared to the full 900 starting PDFs. 

\noindent {\bf PDF4LHC15-30:} A symmetric {\bf Hessian} set with
$N_{\rm eig}=30$. ({Meta-PDF} approach \cite{Gao:2013bia}.) 
This has good precision and is useful
for many experimental needs and when using nuisance parameters

\noindent {\bf PDF4LHC15-100:} A symmetric {\bf Hessian} set with
$N_{\rm eig}=100$ (MC-H) \cite{Carrazza:2015aoa}. This has optimal 
precision if running time is
not a problem or extreme accuracy needed. See Fig. \ref{Fig15} for 
the both Hessian sets compared to the full 900 starting PDFs.


\begin{figure}[]
\vspace{-0.0cm}
\centerline{\includegraphics[width=0.96\textwidth]{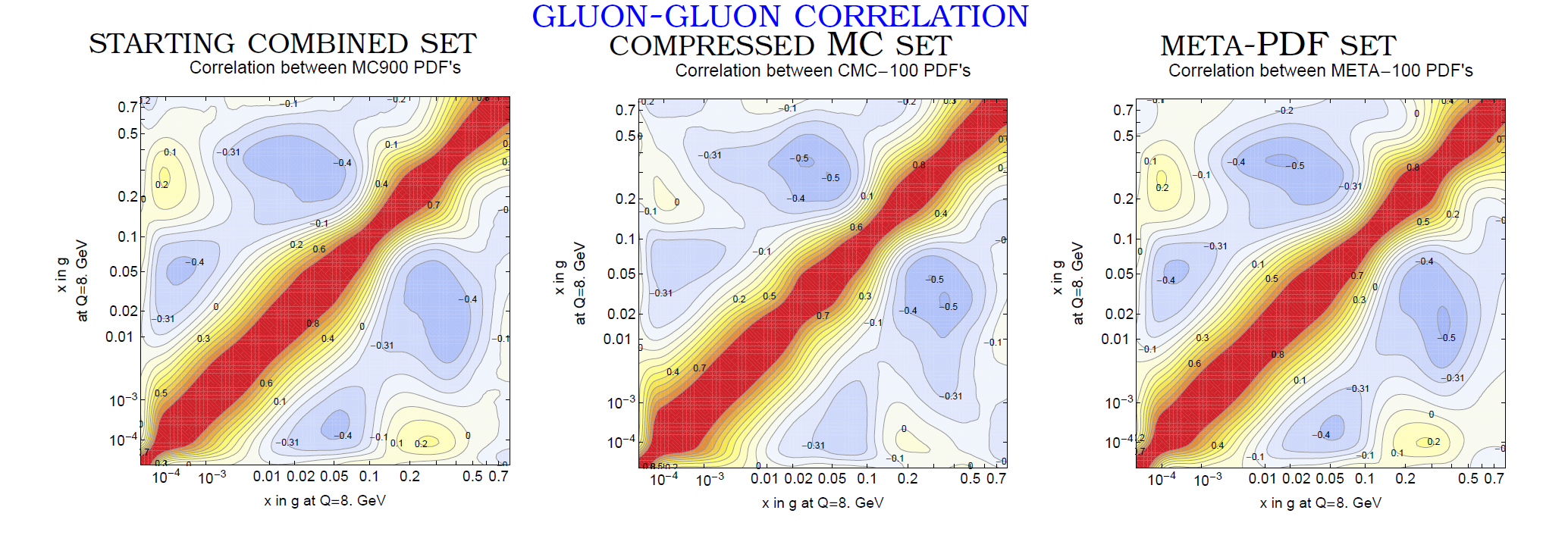}}
\vspace{-0.5cm}
\caption{Comparison of PDF correlations from various means of combination, figures from
\cite{Butterworth:2015oua}}
\vspace{-0.2cm}
\label{Fig17}
\end{figure}

\begin{figure}[]
\vspace{-0.2cm}
\centerline{\includegraphics[width=0.48\textwidth]{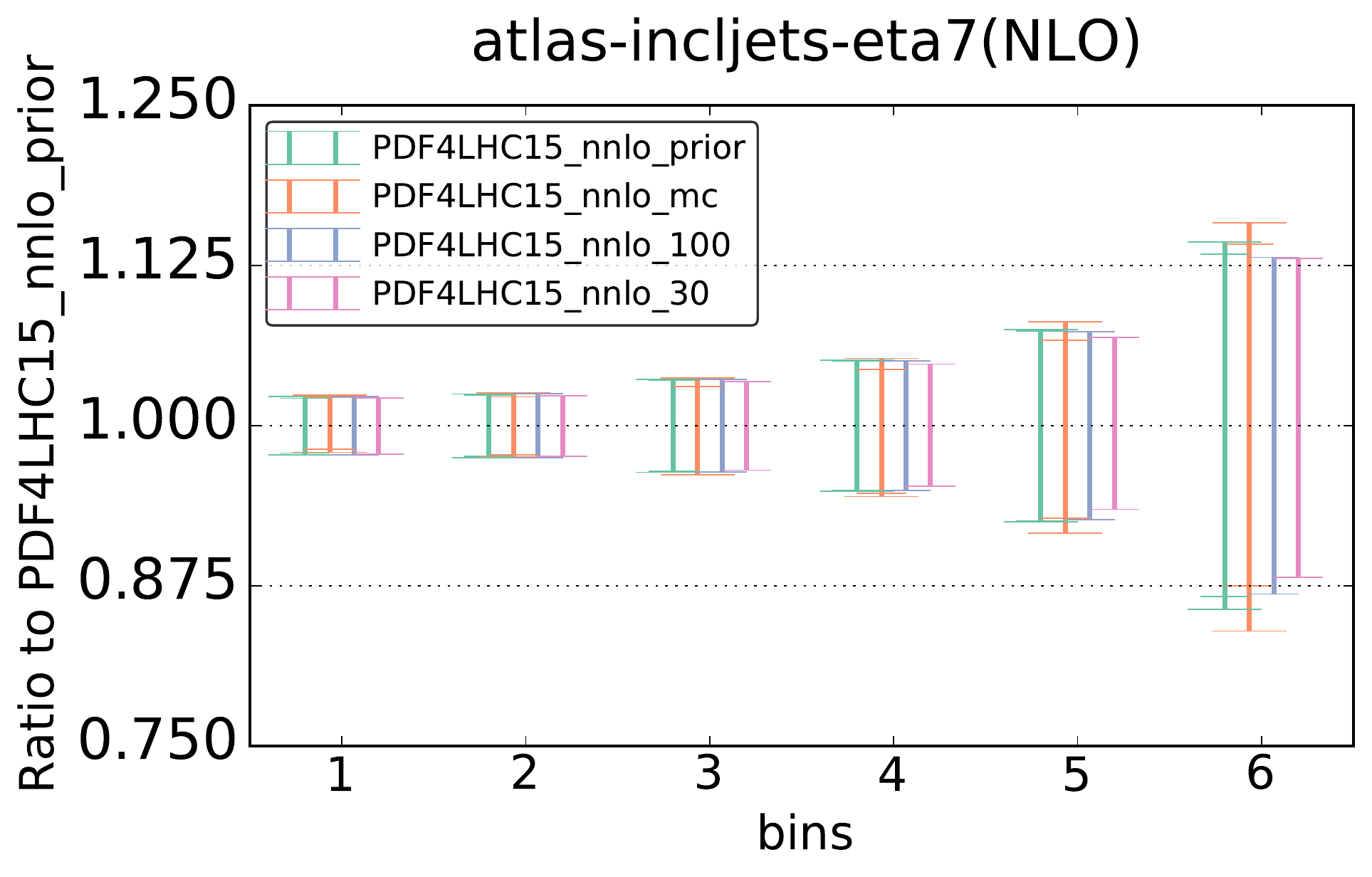}
\includegraphics[width=0.48\textwidth]{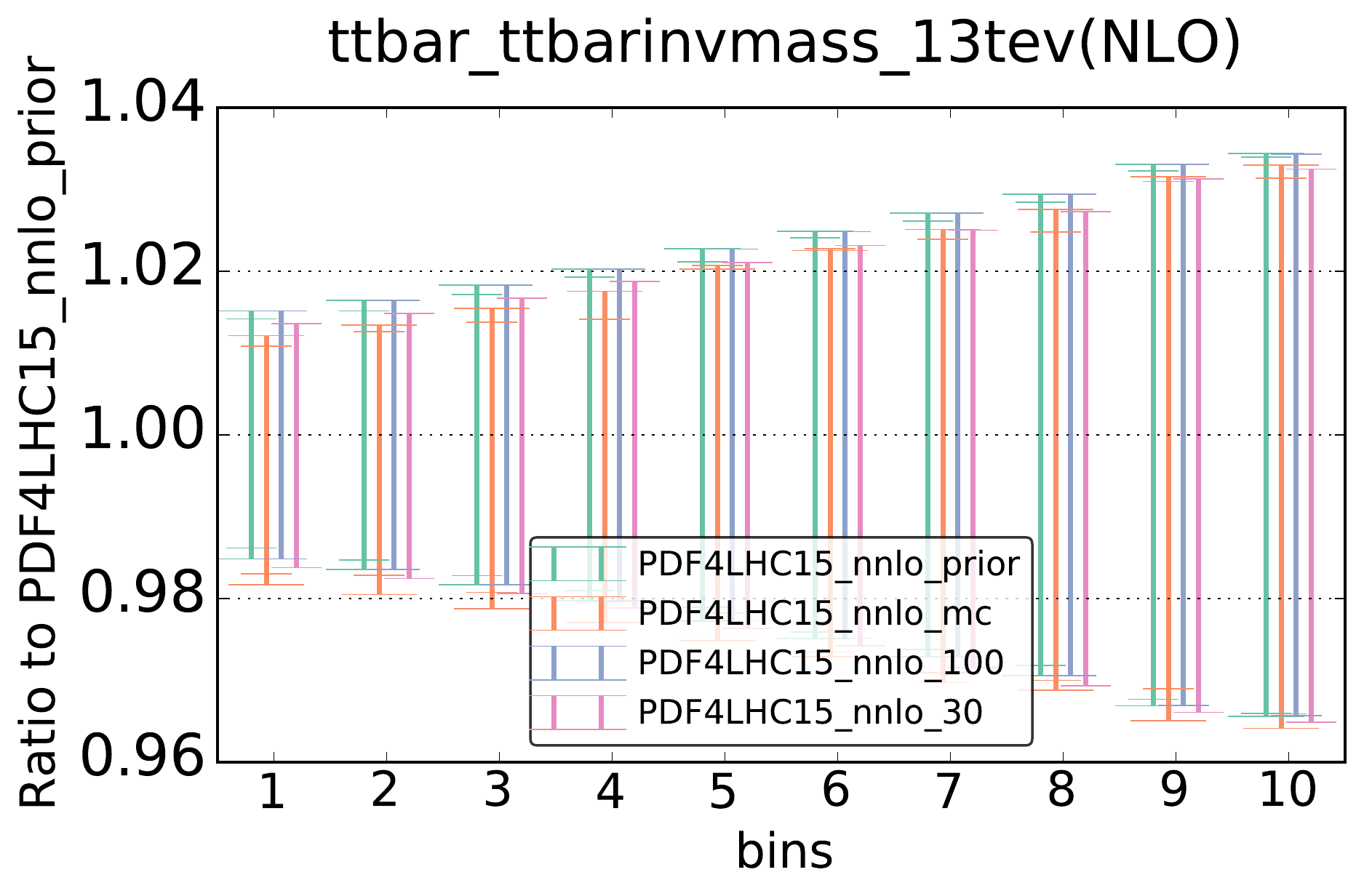}}
\vspace{-0.5cm}
\caption{Examples of differential cross sections using each means of combination, figures from \cite{Butterworth:2015oua}.}
\vspace{-0.2cm}
\label{Fig18}
\end{figure}

PDF correlations are maintained by the compression in all cases. An example is shown in Fig.~\ref{Fig17}. 
The results for cross sections using all the compressed sets for LHC quantities 
work, at worst, quite  well, even in more extreme regions
of kinematics, see Fig.~\ref{Fig18}.

Finally, it is important to note that the PDF4LHC prescription is meant
for assessment of the PDF uncertainty in searches, discovery,
acceptance corrections $\ldots$ (e.g. Higgs, Susy).
When comparing theory predictions to experiment 
in well-determined standard model
processes, e.g. jets, $W,Z$ distributions, it is recommended to use the 
individual PDF sets. Other than the very first measurements at new 
energies processes such as 
top pair cross sections, differential distributions etc will tend to fall into 
the latter category, especially when real precision is reached. 

\vspace{-0.3cm}

\section*{Acknowledgements}

\vspace{-0.1cm}

I would like to thank the members of the MMHT collaboration and of the 
PDF4LHC working group for discussions. This work is
supported partly by the London Centre for Terauniverse Studies (LCTS),
using funding from the European Research Council via the Advanced 
Investigator Grant 267352. I thank the
STFC for support via grant
awards ST/J000515/1 and ST/L000377/1.

\end{document}